\documentclass[reprint,amsmath,amsfonts,amssymb,pra,aps,longbibliography,nofootinbib]{revtex4-1}
\usepackage{graphicx}
\usepackage{siunitx}
\usepackage{braket}
\usepackage{xcolor}

\newcommand{\an}{\hat{a}}
\newcommand{\x}{\hat{x}}
\newcommand{\p}{\hat{p}}
\newcommand{\I}{\hat{I}}
\newcommand{\F}{\hat{F}}
\newcommand{\Ps}{\hat{P}}
\newcommand{\Sq}{\hat{S}}
\newcommand{\R}{\hat{R}}
\newcommand{\CZ}{\hat{C}_Z}
\newcommand{\sech}{\text{sech}}

\begin{document}

	\title{Architecture and noise analysis of continuous-variable quantum gates using two-dimensional cluster states}
	\author{Mikkel V. Larsen}
	\email{mivila@fysik.dtu.dk}
	\author{Jonas S. Neergaard-Nielsen}
	\author{Ulrik L. Andersen}
	\email{ulrik.andersen@fysik.dtu.dk}
	\affiliation{Center for Macroscopic Quantum States (bigQ), Department of Physics, Technical University of Denmark, Fysikvej, 2800 Kgs. Lyngby, Denmark}
	\date{May 27, 2020}

	\begin{abstract}
	Due to its unique scalability potential, continuous variable quantum optics is a promising platform for large scale quantum computing. In particular, very large cluster states with a two-dimensional topology that are suitable for universal quantum computing and quantum simulation can be readily generated in a deterministic manner, and routes towards fault-tolerance via bosonic quantum error-correction are known. In this article we propose a complete measurement-based quantum computing architecture for the implementation of a universal set of gates on the recently generated two-dimensional cluster states \cite{larsen19,asavanant19}. We analyze the performance of the various quantum gates that are executed in these cluster states as well as in other two-dimensional cluster states (the bilayer-square lattice and quad-rail lattice cluster states \cite{alexander16b,menicucci11b}) by estimating and minimizing the associated stochastic noise addition as well as the resulting gate error probability. We compare the four different states and find that, although they all allow for universal computation, the quad-rail lattice cluster state performs better than the other three states which all exhibit similar performance.
	\end{abstract}

	\maketitle

\section{Introduction}
Measurement-based quantum computation (QC) \cite{raussendorf01} on continuous variable (CV) cluster states \cite{menicucci06,gu09}, also known as cluster state computation, shows great potential for scalable quantum information processing. This is due to the simplicity of generating a deterministic and scalable cluster state resource, and the efficiency by which Gaussian gates can be implemented with high-efficiency homodyne detection as already experimentally demonstrated on few-mode cluster states \cite{miwa09,wang10,ukai11a,ukai11b,su13}. The generation of large one-dimensional (1D) cluster states was realized several years ago \cite{yokoyama13,chen14,yoshikawa16}, but for QC at least two dimensions are required---one for encoding and another for computation. There are multiple feasible proposals for the generation of two-dimensional (2D) cluster states \cite{menicucci08,flammia09, menicucci11b, alexander16b,alexander18}, and recently two different 2D cluster states were experimentally realized \cite{larsen19,asavanant19}. A natural question is then, how do the different 2D cluster states compare with regards to their suitability for QC?

Computation schemes for some of the popular 2D cluster states already exist \cite{alexander16a,alexander17, alexander16b,alexander18}. Here, we summarize these schemes and propose new computation schemes for the recently experimentally realized states. Since physical CV cluster states always include noise due to finite squeezing, we furthermore perform a noise analysis of the discussed computation schemes. While similar noise analyses have been done for the 1D dual-rail wire cluster state \cite{alexander14} and the regular 2D square lattice cluster state \cite{menicucci14}, such analysis on experimentally feasible 2D cluster states has not yet, to our knowledge, been carried out. Here, we aim to find the best noise performance for QC with the discussed schemes on each cluster state.

As such, this work is partly a review of existing computation schemes, an introduction to new computation schemes of experimentally realized cluster states, a detailed noise analysis of QC on the different 2D CV cluster states, and a comparison of these. The paper begins with an introduction to the notation and a review of basic concepts in section \ref{sec:2}. In section \ref{sec:3}, we introduce a new computation scheme for the 2D cluster state experimentally realized by us in \cite{larsen19} and perform a noise analysis of this scheme. In section \ref{sec:4}, we describe corresponding computation schemes on three other popular 2D cluster states, namely the quad-rail lattice \cite{alexander16a}, the bilayer square lattice \cite{alexander16b}, and the recently generated cluster state by Asavanant \textit{et al.} \cite{asavanant19}. For each of them, we repeat the same noise analysis as presented in section \ref{sec:3}. In section \ref{sec:5}, we compare the topology and noise performance of the different cluster states and discuss the requirements for universal QC. Finally, we conclude on the results in section \ref{sec:conclusion}. Depending on the reader’s motivation and prior knowledge of cluster state computation, the reader may skip sections and jump to that of interest---we have carefully cross-referenced the sections of this manuscript.

\section{Prerequisite}\label{sec:2}
In this section we review the basic concepts of continuous variable cluster state quantum computation that we will be using in this work. In case the reader is familiar with these concepts, the section can be skipped. 

\subsection{Definitions}
Throughout the paper we assume that $\hbar=1$ and $[\hat{x},\hat{p}]=i$ such that the light field amplitude, $\hat{x}$, (or position) and phase, $\hat{p}$, (or momentum) quadratures can be written as $\x=(\an+\an^\dagger)/\sqrt{2}$ and $\p=-i(\an-\an^\dagger)/\sqrt{2}$, respectively, where $\an$ is the annihilation operator. With these definitions, the variance of the vacuum is 1/2. We will make use of six different unitary operators: The identity operator $\I$, the phase rotation operator $\R(\theta)=e^{-i\theta(\x^2+\p^2)/2}$ (where $\theta$ is the rotation angle) with the Fourier operator $\F=\R(\pi/2)$ as a special case, the squeezing operator $\Sq(s)=e^{i\ln(s)(\x\p+\p\x)/2}$ (where $r=-\ln(s)$ is the standard squeezing parameter), the shear operator $\Ps(p)=e^{ip\x^2/2}$ (where $p$ is a shearing parameter), the controlled-Z operator $\CZ(g)=e^{ig\x\otimes\x}$ (where $g$ is the coupling coefficient), and the balanced beam splitter operator $\hat{B}=e^{-i\pi(\x\otimes\p-\p\otimes\x)/4}$. Each of these operators are Gaussian and can be described by symplectic matrices representing the evolution of the quadrature operators, arranged in a vector $(\x_1,\cdots,\x_n,\p_1,\cdots,\p_n)^T$ for $n$ modes, in the Heisenberg picture:
\begin{equation*}
	\mathbf{I}=\begin{pmatrix}	1 & 0 \\ 0 & 1	\end{pmatrix}\quad,\quad
	\mathbf{R}=\begin{pmatrix} \cos\theta & \sin\theta \\ -\sin\theta & \cos\theta	\end{pmatrix}\quad,\quad
\end{equation*}
\begin{equation*}
	\mathbf{S}=\begin{pmatrix}	\frac{1}{s} & 0 \\ 0 & s	\end{pmatrix}\quad,\quad
	\mathbf{P}=\begin{pmatrix} 1 & 0 \\ p & 1	\end{pmatrix}\quad,\quad
\end{equation*}
\begin{equation*}
	\mathbf{C}_\mathbf{Z}=\begin{pmatrix}
	1 & 0 & 0 & 0\\
	0 & 1 & 0 & 0\\
	0 & g & 1 & 0\\
	g & 0 & 0 & 1
	\end{pmatrix}\quad,
\end{equation*}
and
\begin{equation}\label{eq2:B}
	\mathbf{B}=\frac{1}{\sqrt{2}}\begin{pmatrix}
	1 & -1 & 0 & 0\\
	1 & 1 & 0 & 0\\
	0 & 0 & 1 & -1\\
	0 & 0 & 1 & 1
	\end{pmatrix}\quad,
\end{equation}
for $\I$, $\R(\theta)$, $\Sq(s)$, $\Ps(p)$, $\CZ(g)$ and $\hat{B}$, respectively.

To allow for quantum error correction, in this paper we consider the encoding of qubits in bosonic modes of computation, $\ket{\psi_\text{in}}$. Numerous different qubit encodings have been proposed such as encoding in cat states \cite{cochrane99,lund08} and binomial states \cite{michael16,albert18} but here we will consider the efficient Gottesman-Kitaev-Preskill (GKP) encoding \cite{gottesman01}. For these codes, a qubit is encoded on a square lattice in phase-space in a way that allows for the suppression of relevant errors (such as loss) to a certain extend. To combat residual qubit errors, the GKP code must be concatenated with another qubit error correction code such as the 7-qubit Steane code \cite{steane03} or Knill's $C_4/C_6$ code \cite{knill05}. GKP-encoding is the only known bosonic code for which a universal Gaussian gate set allows logic Clifford computation and error correction of encoded qubits. A logic single-mode Clifford gate set is realized by the Gaussian gate set $\lbrace\I,\F,\Ps(1)\rbrace$ together with displacements, while two-mode gates are enabled by the $\CZ(1)$-gate. A non-Clifford gate completes the universal encoded qubit gate set. While the non-Clifford gate requires challenging non-Gaussian transformations, practical proposals do exists which are further discussed in section \ref{sec:6}. Here in section \ref{sec:2}, and in section \ref{sec:3} and \ref{sec:4}, we will focus on the implementations of the Gaussian gates $\I$, $\F$, $\Ps(1)$ and $\CZ(1)$.

Gaussian gates are implemented on a cluster state by quadrature measurement of each mode in different bases rotated by $\theta$ with respect to the $\x$-quadrature, i.e. measuring $\x(\theta)=\x\cos\theta+\p\sin\theta$. In this paper we use the Heisenberg picture, in which we simulate the evolution of the quadrature operators and where the noise contributions simply appear as additive Gaussian noise terms. In the following we consider the generalized teleportation circuit as an example, which as well plays an important role in the quantum computation schemes presented in section \ref{sec:3} and \ref{sec:4}.

\begin{figure}
	\includegraphics[width=\linewidth]{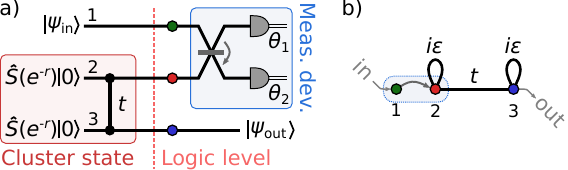}
	\caption{\label{fig2:gentele}(a) Generalized teleportation circuit implementing single mode Gaussian operation on the input state $\ket{\psi_\text{in}}$ by joint measurement of the input state with one mode of an ancillary approximate cluster state using a two-mode measurement device (meas. dev.). The measurement device consists of a beam splitter, with the arrow pointing from the first to the second mode in Eq.~(\ref{eq2:B}), followed by homodyne detection measuring in bases $\x_i(\theta_i)=\x_i\cos\theta_i+\p_i\sin\theta_i$. (b) Short graphical notation of the circuit in (a) used in this paper, with the nodes representing the modes in (a). Here, $\varepsilon=e^{-2r}$. The node colors have no physical meaning and are only used to identify the modes in (a). The graphical notation can be thought of as a ``snap-shot'' of the logic level in (a) where the computation takes place.}
\end{figure}

\subsection{Generalized teleportation}\label{sec2:gentele}
An arbitrary Gaussian transformation on a single bosonic mode can be realized by means of the generalized teleportation circuit, as diagrammatically depicted in Fig.~\ref{fig2:gentele}a. Here, the term \textit{generalized} teleportation is in terms of the generalized measurement bases, where different gates are implemented on a teleported state depending on the basis setting. Generalized teleportation consists of an input state, an entangled multi-mode ancillary state, and a measurement device. In conventional single-mode teleportation, the ancillary entangled state is a two-mode squeezed state \cite{furusawa98}, while traditionally for MBQC we consider a cluster state. The two-mode squeezed state and two-mode cluster state are equivalent (up to a phase rotation), but since well-developed theoretical tools exist for cluster states in the language of graphical calculus \cite{menicucci11a}, here we focus on cluster states. In practice the ancillary entangled state is an approximate cluster state composed of finitely momentum-squeezed states that are entangled by a controlled-Z gate of weight $t$ \cite{gu09}. In Fig.~\ref{fig2:gentele}a, the momentum variance is $\varepsilon/2$ where $\varepsilon=e^{-2r}$ with $r$ being the squeezing parameter. To implement a gate, a joint measurement is performed on the input state and one mode of the cluster state using the measurement device marked on Fig.~\ref{fig2:gentele}a consisting of a beam splitter and two homodyne detectors. The resulting transformation of the quadratures in the Heisenberg picture is
\begin{equation}\label{eq2:trans}
	\begin{pmatrix}
		\x'_3\\\p'_3
	\end{pmatrix}=\mathbf{G}
	\begin{pmatrix}
		\x_1\\\p_1
	\end{pmatrix}+\mathbf{N}
	\begin{pmatrix}
		\p_2\\\p_3
	\end{pmatrix}+\mathbf{D}
	\begin{pmatrix}
		m_1\\m_2
	\end{pmatrix}\;.
\end{equation}
Here, the first term represents the implemented Gaussian gate with $\mathbf{G}$ corresponding to the gate symplectic matrix. The second term, with \textbf{N} being a matrix, represents noise added to the quadratures due to finite squeezing in the cluster state; it vanishes in the infinite squeezing limit as $\braket{\p_{2,3}}=0$ and $\text{Var}(\p_{2,3})=e^{-2r}/2\rightarrow0$ for $r\rightarrow\infty$. This term represents the \textit{gate noise}. The last term, with \textbf{D} being a matrix, is the computational by-product in the form of a displacement, where $m_1$ and $m_2$ are the measurement outcomes of mode 1 and 2, respectively. \textbf{G}, \textbf{N}, and \textbf{D} are each described in detail in the following subsections.

When considering multi-mode computing schemes with large cluster states, the circuit model in Fig.~\ref{fig2:gentele}a becomes tedious. Instead it is customary to use a graph notation as illustrated in Fig.~\ref{fig2:gentele}b, where the cluster state is represented by its corresponding graph with imaginary self-loops indicating the finite squeezing of the cluster state modes \cite{menicucci11a}, the beam splitter of the measurement device is represented by an arrow, and the input state is represented by a free node. For the schemes presented in this work, we assume that all cluster state modes are equally squeezed.
Hence, we will omit the identical $i\varepsilon$ self-loops on the cluster state nodes---they are always there, and only vanish in the non-physical infinite squeezing limit. Finally, we will define the \textit{logic level} as being the level in the circuit diagram where the computation takes place, i.e. after the cluster state generation where the input state appears, and before the measurement device for computation. The logic level is marked on Fig.~\ref{fig2:gentele}a, and the graph notation in Fig.~\ref{fig2:gentele}b is a ``snap-shot'' of this logic level with the arrow indicating the subsequent beam splitter operation of the measurement device. An alternative formulation is to use macronodes as in \cite{alexander14,alexander16b}, where, instead of joint measurements of localized modes in the logic level, one considers single-mode measurements of distributed modes. In Fig.~\ref{fig2:gentele}a this macronode formulation corresponds to locating the logic level right after the beam-splitter transformation, and keeping in mind, that in the logic level the mode under computation is distributed between mode 1 and 2 as $(\an_1+\an_2)/\sqrt{2}$.

In the following we describe each term of Eq.~\eqref{eq2:trans} in more detail:

\subsubsection{Gate}\label{sec2:gate}
The implemented gate in Eq.~(\ref{eq2:trans}) depends on the measurement bases of the two quadrature measurements (defined as $\theta_1$ and $\theta_2$) as
\begin{equation*}
	\mathbf{G}=\frac{1}{\sin\theta_-}\begin{pmatrix}
		\frac{1}{t}\cos\theta_++\frac{1}{t}\cos\theta_- & \frac{1}{t}\sin\theta_+\\
		-t\sin\theta_+ & t\cos\theta_+-t\cos\theta_-
	\end{pmatrix}\;,
\end{equation*}
where $\theta_\pm=\theta_1\pm\theta_2$, and corresponds to the operation \cite{ukai10}
\begin{equation}\label{eq2:gate}
	\Sq(t)\R\left(\frac{\theta_+}{2}\right)\Sq\left(\tan\frac{\theta_-}{2}\right)\R\left(\frac{\theta_+}{2}\right)\;.
\end{equation}
By implementing such operation twice---corresponding to two consecutive runs of the teleportation circuit in Fig~\ref{fig2:gentele}a---it is possible to induce an arbitrary single mode Gaussian transformation \cite{alexander14}, $\textbf{U}(\x,\p)^T+\textbf{c}$, where
$\textbf{U}=\textbf{G}_2\textbf{G}_1$ while the displacement $\textbf{c}$ is ubiquitous and simply implemented by shifting the measurement result and updating the bases of the subsequent measurements \cite{menicucci06,gu09}. However, in the following we consider only the subset of single-mode Gaussian transformation that is required for GKP state computation, namely the set $\lbrace\I,\F,\Ps(1)\rbrace$.

The identity operator, $\hat{I}$, can be executed in a single computation step of the teleportation circuit in Fig.~\ref{fig2:gentele}a by choosing 
\begin{equation*}
    (\theta_+,\theta_-)_I=(0,2\arctan 1/t)\;,
\end{equation*}
as easily seen from Eq.~\eqref{eq2:gate}. For the Fourier gate, $\F$, and the $\Ps(1)$-gate two computation steps are necessary: If we let the output state $\ket\psi_\text{out}$ of the first computation step in Fig.~\ref{fig2:gentele}a, with bases $(\theta_{+1},\theta_{-1})$, be the input state on a second identical circuit, but with bases $(\theta_{+2},\theta_{-2})$, the gate
\begin{equation*}\begin{aligned}
	&\Sq(t)\R\left(\frac{\theta_{+2}}{2}\right)\Sq\left(\tan\frac{\theta_{-2}}{2}\right)\R\left(\frac{\theta_{+2}}{2}\right)\times\\
	&\quad\quad\quad\Sq(t)\R\left(\frac{\theta_{+1}}{2}\right)\Sq\left(\tan\frac{\theta_{-1}}{2}\right)\R\left(\frac{\theta_{+1}}{2}\right)
\end{aligned}\end{equation*}
is implemented. Choosing
\begin{equation*}
	\left(\theta_{+1},\theta_{-1},\theta_{+2},\theta_{-2}\right)_F=\left(\frac{\pi}{2},\frac{\pi}{2},0,2\arctan\frac{1}{t^2}\right)
\end{equation*}
implements the $\F$-gate, while
\begin{equation*}
	\left(\theta_{+1},\theta_{-1},\theta_{+2},\theta_{-2}\right)_P=\left(\arctan2,-\arctan2,\frac{\pi}{2},\frac{\pi}{2}\right)
\end{equation*}
implements the $\Ps(1)$-gate. To implement the two-mode $\CZ$-gate a scheme with at least two input modes is required. This will be discussed in section \ref{sec:3} and \ref{sec:4}, where we find that also two teleportation steps are necessary.

\subsubsection{Gate noise}\label{sec2:gate_noise}
The second term in Eq.~(\ref{eq2:trans}) represents the gate noise and is governed by the matrix
\begin{equation*}
	\mathbf{N}=\begin{pmatrix}
		-\frac{1}{t}&0\\0&1
	\end{pmatrix}\;,
\end{equation*}
which is independent on the measurement bases. However, for gates realized in two steps, as the $\F$- and $\Ps(1)$-gate described above, the gate noise of the first step enters the gate of the second step, leading to the final gate noise that depends on the bases of the second computation step. For two concatenated circuits of the type in Fig.~\ref{fig2:gentele}a, with the cluster state of the second circuit being denoted mode 4 and 5, the combined gate noise becomes
\begin{equation}\label{eq2:2step_gate_noise}
	\mathbf{G}_2\mathbf{N}_1\begin{pmatrix}
	\p_2\\\p_3
	\end{pmatrix}+\mathbf{N}_2\begin{pmatrix}
	\p_4\\\p_5
	\end{pmatrix}\equiv\mathbf{N}\begin{pmatrix}
	\p_2\\\p_3\\\p_4\\\p_5
	\end{pmatrix}\;,
\end{equation}
where $\mathbf{N}_1(\p_2,\p_3)^T$ is the gate noise of the first step, and $\mathbf{G}_2$ and $\mathbf{N}_2(\p_4,\p_5)^T$ is the gate symplectic matrix and gate noise of the second step. Here, the combined gate noise matrix $\mathbf{N}$ is a $2\times4$ matrix. Note, that in the following sections III and IV, \textbf{N} is in general the combined gate noise matrix of an implemented gate in one or more teleportation steps, with the number of columns equal to the number of quadratures in the output mode(s), and the number of rows equal to the number of ancillary cluster state modes involved in the implemented gate.

Assuming that all cluster state modes in Eq.~\eqref{eq2:2step_gate_noise} are equally squeezed, $\text{Var}(\hat{p}_l)=\varepsilon/2$ for $l=2,3,4,5$, the gate noise variance amounts to $\sum_{j=1}^4 N_{ij}^2\varepsilon/2$ for $i=1,2$. From this expression, it is clear that the gate noise can be reduced by both increasing the degree of squeezing of the cluster state modes (reducing $\varepsilon/2$) and by minimizing the sums $\sum_{j=1}^4 N_{ij}^2$, $i=1,2$. We will refer to these two sums as the \textit{quadrature noise factors}, one for each quadrature. In the schemes for gate implementation presented in section \ref{sec:3} and \ref{sec:4}, the focus is on optimizing the quadrature noise factors for the gate set $\lbrace\I,\F,\Ps(1),\CZ(1)\rbrace$ in order to minimize the probability of inducing errors on the GKP-encoded qubit as discussed in section \ref{sec2:errc}.

\subsubsection{Displacement by-product}
The displacement matrix in Eq.~(\ref{eq2:trans}) reads
\begin{equation*}
	\mathbf{D}=\frac{\sqrt{2}}{\sin\theta_-}\begin{pmatrix}
	-\frac{1}{t}\cos\theta_2&-\frac{1}{t}\cos\theta_1\\t\sin\theta_2&t\sin\theta_1
	\end{pmatrix}\;,
\end{equation*}
and so, since we know the measurement bases, $(\theta_1,\theta_2)$, together with the measurement outcome, $(m_1,m_2)$, the amount of displacement of a single computation step is known. By keeping track of the displacements in each computation step, the displacement can be accounted for in the following steps by updating the measurement bases and results (i.e. \textit{feed-forward}) \cite{gu09}. When all gates are Gaussian, the displacement by-product can be compensated for in the measurement result of the final output state $\ket{\psi_\text{out}}$, known as \textit{Gaussian parallelism} \cite{menicucci06}.

In this work we will ignore the displacement by-product as it has no effect on the gate noise performance. However, for the actual experimental implementations of the schemes discussed in this work, the compensation for the displacement is important.

\subsubsection{Wigner function representation}
To understand the effect of the quadrature transformation and gate noise in Eq.~\eqref{eq2:trans}, it is useful to analyse the generalized teleportation in the Wigner function representation. This does not add anything new, but considering the gate noise from a different perspective helps to understand it. Here, the Wigner function of the input state to the teleportation circuit in Fig.~\ref{fig2:gentele}, $\ket{\psi_\text{in}}_1\Sq(e^{-r})\ket{0}_2\Sq(e^{-r})\ket{0}_3$, is 
\begin{equation*}
	W_\text{in}(x_1,p_1)G_{1/\varepsilon}(x_2)G_\varepsilon(p_2)G_{1/\varepsilon}(x_3)G_\varepsilon(p_3)\;,
\end{equation*}
where $G_\delta$ is a normalized Gaussian function of variance $\delta/2$, and $W_\text{in}$ is the Wigner function corresponding to $\ket{\psi_\text{in}}_1$, but not necessarily of a pure state. After the quadrature transformation in the generalized teleportation circuit and measurement of $\x_1(\theta_1)$ and $\x_2(\theta_2)$, the output Wigner function corresponding to $\ket{\psi_\text{out}}$ on mode 3 in Fig.~\ref{fig2:gentele} becomes
\begin{widetext}
	\begin{equation}\label{eq2:Wigner}
		W_\text{out}(x_3,p_3)=\mathcal{N}G_{1/\varepsilon}(x_3)\int\text{d}\eta_2\,G_{\varepsilon}\left(\eta_2\right) G_{t^2/\varepsilon}\left(p_3-\eta_2\right)\int\text{d}\eta_1\,G_{\varepsilon/t^2}\left(\eta_1\right)	W_\text{in}\left(\textbf{G}^{-1}\begin{pmatrix}
			x_3-d_x-\eta_1\\p_3-d_p-\eta_2
		\end{pmatrix}\right)
		\;,
	\end{equation}
\end{widetext}
where $\mathcal{N}$ is a normalization factor depending on the basis setting $(\theta_1,\theta_2)$ and measurement outcome $(m_1,m_2)$, and  $(d_x,d_p)^T=\textbf{D}(m_1,m_2)^T$ subtracted from the $(x_3,p_3)^T$ argument in $W_\text{in}$ corresponds to the displacement by-product. Here, each index in the vector argument of $W_\text{in}$ should be understood as the two arguments in $W_\text{in}(\cdot,\cdot)$. It is clear from the expression that the input state undergoes an operation of symplectic matrix $\textbf{G}$ by the transformation of its Wigner function arguments by $\textbf{G}^{-1}$. This corresponds to the implemented gate. The gate noise $\textbf{N}(\p_2,\p_3)^T$ with variance $(\varepsilon/t^2,\varepsilon)^T/2$ is seen to become a Gaussian convolution of the Wigner function with corresponding variance after applying the gate $\textbf{G}$. Finally, the Wigner function is subjected to a Gaussian envelope in both quadratures: One of variance $t^2/2\varepsilon$ in the $\p$-quadrature after convolution in $\x$-quadrature with $G_{\varepsilon/t^2}$, followed by one of variance $1/2\varepsilon$ in $\x$-quadrature after convolution in $\p$-quadrature with $G_\varepsilon$. These envelopes in each quadratures are the result of convolutions in orthogonal quadratures, as the quadratures are related by the Fourrier transform: Convoluting a Wigner function in $\x$-quadrature with $G_\delta$ leads to an envelope of the Wigner function in the $\p$-quadrature of $G_{1/\delta}(p)$, and vice versa.

Two limits of Eq.~\eqref{eq2:Wigner} are interesting: In the ideal infinite squeezing limit, the convolution functions $G_{\varepsilon/t^2}$ and $G_{\varepsilon}$ becomes delta functions, while the Gaussian envelopes becomes infinitely broad, and so $W_\text{out}(x_3+d_x,p_3+d_p)^T=W_\text{in}\left(\textbf{G}^{-1}(x_3,p_3)^T\right)$. In the limit $t=0$, where we expect no information to transfer from $W_\text{in}$ to $W_\text{out}$, the envelope on $W_\text{in}\left(\textbf{G}^{-1}(x_3,p_3)^T\right)$ becomes a delta function in $\p$-quadrature, which is then convoluted by $G_\varepsilon$, while the $\x$-quadrature is convoluted by an infinitely broad Gaussian followed by an envelope of $G_{1/\varepsilon}$. As a result, for $t=0$, $W_\text{out}(x_3,p_3)=G_{1/\varepsilon}(x_3)G_{\varepsilon}(p_3)$ corresponding to the initial input squeezed state in the cluster state as expected.

\subsection{Error correction}\label{sec2:errc}
It is now clear that cluster state quantum computation will inevitably suffer from gate noise that will accumulate throughout the computation. To avoid noise accumulation, quadrature error correction is required in between every implemented gate. For this purpose, symmetric GKP states are particularly useful, not only as the qubit but also as ancillaries for error correction. GKP states have been thoroughly reviewed in several places \cite{tzitrin20,terhal20}, while here it is reviewed in terms of MBQC focusing on the added gate noise caused by finite squeezing in the cluster states. The GKP state Wigner function,  $W_\text{in}(\textbf{x})$ (with $\textbf{x}=(x_1,\cdots,x_n,p_1,\cdots,p_n)^T$ for a $n$-mode state), consists of delta functions arranged on a square lattice in phase space of each mode with a lattice constant of $\sqrt{\pi}$, and its qubit eigenstates of the Pauli-Z and -X operators are $\ket{j_L}_{X,Z}=\sum_{i\in\mathbb{Z}}\ket{(2i+j)\sqrt{\pi}}_{x,p}$ in the quadrature eigenstate basis, $\ket{s}_x$ and $\ket{s}_p$ \cite{gottesman01}.

As a result of the execution of a $n$-mode gate $\textbf{G}$ in one or more computation steps, the gate noise $\textbf{N}(\p_{c1},\p_{c2},\cdots)^T$ (where $p_{ci}$ are ancillary cluster state modes) leads to a broadening of the GKP delta functions into Gaussian functions of variances in $\boldsymbol{\sigma^2}=\text{Var}\lbrace\textbf{N}(\p_{c1},\p_{c2},\cdots)^T\rbrace$ in the $2n$ quadratures ($\boldsymbol{\sigma^2}$ is a $2n$ vector). Furthermore, as the ideal GKP-encoding with vanishing variance (represented by delta functions) is non-physical, we instead consider the physical approximate GKP-states in $W_\text{in}(\textbf{x})$ with delta functions replaced by symmetric Gaussian functions of identical variance, $\delta$, in $\x$- and $\p$-quadrature. The quadrature variance of the Gaussian spikes in the approximate GKP-state after the implementation of a noisy gate is then
\begin{equation}\label{eq2:deltam}
	\boldsymbol{\delta'}=\text{Var}\lbrace\textbf{G}\hat{\textbf{x}}_{\boldsymbol{\delta}}\rbrace+\boldsymbol{\sigma^2}\;,
\end{equation}
where $\hat{\textbf{x}}$ is decomposed into a sum,  $\hat{\textbf{x}}_\textbf{0}+\hat{\textbf{x}}_{\boldsymbol{\delta}}$, where $\hat{\textbf{x}}_\textbf{0}$ and $\hat{\textbf{x}}_{\boldsymbol{\delta}}$ are the centers and variance of the GKP spikes, respectively. Note that for ideal GKP states $\hat{\textbf{x}}=\hat{\textbf{x}}_\textbf{0}$. As examples, $\text{Var}\lbrace\textbf{G}\textbf{x}_{\boldsymbol{\delta}}\rbrace=(\delta,\delta)^T$ for the $\I$- and $\F$-gate, $\text{Var}\lbrace\textbf{G}\textbf{x}_{\boldsymbol{\delta}}\rbrace=(\delta,2\delta)^T$ for the $\Ps(1)$-gate, and $\text{Var}\lbrace\textbf{G}\textbf{x}_{\boldsymbol{\delta}}\rbrace=(\delta,\delta,2\delta,2\delta)^T$ for the two-mode $\CZ(1)$-gate.

To avoid gate noise accumulating on the GKP-encoded qubit state, after every implemented $\I$-, $\F$-, $\Ps(1)$- and $\CZ(1)$-gate, we measure the quadratures $\x$ mod $\sqrt{\pi}$ and $\p$ mod $\sqrt{\pi}$ using ancillary GKP-states, and perform quadrature error correction by displacing back the state depending on the measurement outcome:
\begin{equation}\label{eq:ercCircuit}
	\includegraphics[width=0.85\linewidth]{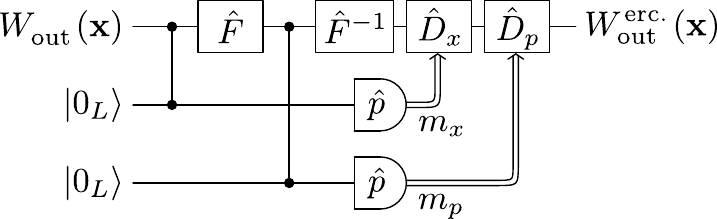},
\end{equation}
where $\ket{0_L}$ are approximate GKP-states with spike variances in both quadratures of $\delta$. Note that while this circuit illustrates the correction algorithm for a single mode of $W_\text{out}$, similar circuits are required for each other mode. After the two measurements with outcome $m_x$ and $m_p$, the encoded qubit is projected into a ``fresh'' GKP-state, but displaced in $\x$- and $\p$-quadrature depending on the values of $m_x$ and $m_p$: If $m_{x(p)}$ mod $\sqrt{\pi}$ is smaller than $\sqrt{\pi}/2$, the encoded state is displaced back by $m_{x(p)}$ mod $\sqrt{\pi}$ in $\x(\p)$, while if it is larger than $\sqrt{\pi}/2$, the encoded states is displaced forward by $\sqrt{\pi}-(m_{x(p)}\text{ mod }\sqrt{\pi})$, and so we obtain an error corrected version of $W_\text{out}(\textbf{x})\rightarrow W_\text{out}^\text{erc.}(\textbf{x})$, which is then the input to the next gate. The possible values of $m_x$ and $m_p$ are Gaussian distributed with variance $\delta'_i+\delta$, where $\delta'_i$ in $\boldsymbol{\delta'}$ of Eq.~\eqref{eq2:deltam} is the corresponding spike quadrature variance of the encoded state after gate implementation, $W_\text{out}$, and $\delta$ is the spike variance of the $\ket{0_L}$ ancillary states. As a result, for large $\delta_i'$ and/or $\delta$, there is a risk of measuring a GKP spike closer to its neighbouring spikes of the orthogonal qubit state, i.e. outside the bin range $[x_0-\sqrt{\pi}/2;x_0+\sqrt{\pi}/2]$ where $x_0$ is the spike center, and thereby inducing a qubit error when ``correcting'' the state by displacing it in the wrong direction. The combined probability of displacing a $n$-mode encoded state with $2n$ quadrature corrections in the wrong direction is shown to be \cite{menicucci14}
\begin{equation}\label{eq2:Perr}
	P_\text{err.}(\boldsymbol{\delta'},\delta)=1-\prod_{i=1}^{2n}\text{erf}\left(\frac{\sqrt{\pi}}{2\sqrt{2(\delta_i'+\delta)}}\right)\;,
\end{equation}
where each factor in the product term is the probability of a successful quadrature correction. It is important to mention that $P_\text{err.}$ is not a true qubit error probability, as it does not account for the probability of measuring a spike at its neighbours neighbour bin range, $[x_0\pm3\sqrt{\pi}/2;x_0\pm5\sqrt{\pi}/2]$, which leads to a $2\sqrt{\pi}$ displacement of the GKP state when corrected, and thereby not a qubit error although it is an error. This leads for example to $P_\text{err.}\rightarrow 1$ for large $\delta_i'+\delta$ while the actual error probability should be 1/2. Furthermore, Eq.~\eqref{eq2:Perr} does not account for the overall envelope on the spikes of the GKP-state, and for the fact that the error probability is qubit-dependent: Displacing the $\p$-quadrature by $\sqrt{\pi}$ leads to an error on $\ket{+_L}$, but no error on $\ket{0_L}$. Therefore, for a true estimation of the qubit error probability, we need to take these effects into account. However, despite these issues, in this work (as in \cite{menicucci14}) we will use $P_\text{err.}$ as a figure of merit as it constitutes a good approximation to the actual error probability for reasonably large squeezing levels in which $\delta_i'+\delta$ is small enough for $2\sqrt{\pi}$ (or larger) displacements to be neglected during quadrature corrections.        

Since the two-mode $\CZ(1)$-gate requires four quadrature corrections, while the $\I$-, $\F$- and $\Ps(1)$-gate only requires two, the error probability after the $\CZ(1)$-gate is in general larger. In the schemes presented in section \ref{sec:3} and \ref{sec:4}, when possible, we search for a basis setting for the $\CZ(1)$-gate that minimizes $P_\text{err.}$.

\section{Double bilayer square lattice}\label{sec:3}
\begin{figure*}
	\includegraphics[width=\linewidth]{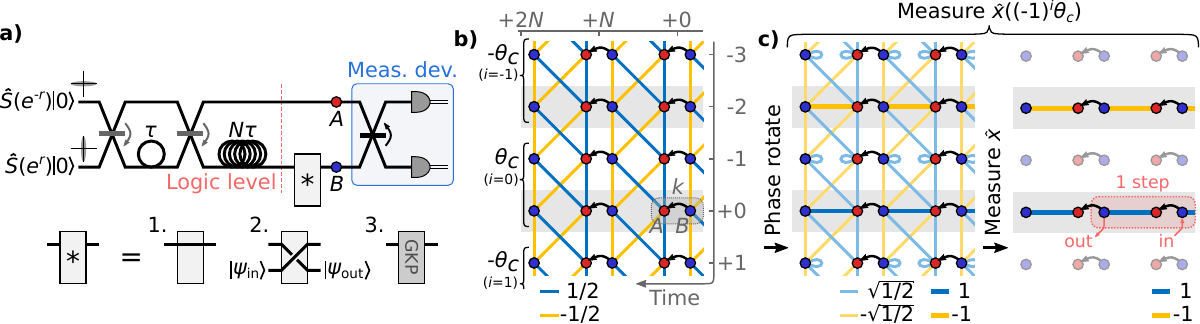}
	\caption{\label{fig3:DBSL}(a) Temporal encoded DBSL cluster state generation and computation setup. While the DBSL cluster state is generated after the third beam splitter, for computation we consider the cluster state in the marked logic level, and the third beam splitter as part of a joint measurement device for computation. The cluster state temporal mode duration $\tau$ is defined by the short delay line. The device marked by asterisk is an identity gate when implementing gates (1.), an optical switch when switching in a state for computation (2.), or the circuit in Eq.~(\ref{eq:ercCircuit}) for correcting gate noise after each implemented gate (3.). (b) Cluster state in the logic level: A dual-rail wire coiled up by the $N\tau$ long delay, leading to a cylinder with $N$ temporal modes in the circumference---the temporal mode indices are written in grey. Computation is performed in wires including the modes in the marked grey areas, while modes in between the grey areas are control modes. Measuring the control modes in an alternating basis of $(-1)^i\theta_c$ induces edges between the wire modes as shown in (c), allowing single mode computation in each $N/2$ wire with one computation step marked in red. Measuring a control mode in a different basis creates coupling between the neighbouring two wires as shown in Fig.~\ref{fig3:CZ} allowing multi-mode gates. The edge weights shown here is in the limit of infinite squeezing and $\theta_c=\pi/4$. For finite squeezing the edge weights are multiplied by $\tanh(2r)$ while self-loops of $i\sech(2r)$ are present on each cluster state mode. For easy comparison, the experimental setup, logic cluster state and its projection into wires are shown in appendix \ref{app:C} together with the schemes considered in section \ref{sec:4}.}
\end{figure*}
Having discussed the general concept of CV quantum computation and the associated error analysis, we are now equipped with the relevant tools to rigorously analyze the performance of cluster state computation based on different types of cluster states. In this section, we will consider the \textit{double bilayer square lattice} (DBSL) cluster state while in the following section \ref{sec:4} we will consider three other known clusters states.

The cylindrical 2D cluster state produced in Ref. \cite{larsen19}, can be straightforwardly projected into a universal DBSL cluster which will be analysed in the following. The cylindrical DBSL cluster state with a 2D topology of Ref.~\cite{larsen19} (corresponding to a cylindrical $\mathcal{H}$-graph state) was generated by ``coiling up'' a 1D cluster state (a dual-rail wire \cite{yokoyama13}) of temporal mode duration $\tau$ using a $N\tau$ long delay line, and interfering it with itself---the generation setup is summarized in Fig.~\ref{fig3:DBSL}a. As this $\mathcal{H}$-graph state is self-inverse and bipartite for even $N$, it is transformed into a DBSL cluster state through $\pi/4$ phase rotations of all modes. Since this transformation simply corresponds to a redefinition of the quadratures, the DBSL $\mathcal{H}$-graph state and the corresponding cluster state are equivalent \cite{larsen19,menicucci11a}. In the following we will therefore only consider the DBSL cluster state. 

It is also important to note that in Ref. \cite{larsen19}, it was shown that the DBSL cluster state can be projected into a regular square lattice cluster state which is known to be a universal resource for quantum computing. However, due to the resulting low edge weights of this square lattice, this approach is rather inefficient and leads to unnecessary large gate noise. In the following, we present a more efficient computation scheme of the DBSL and quantify it by a gate noise analysis.

\subsection{Efficient computation scheme}\label{sec:3A}
Similar to the generalized teleportation scheme in section \ref{sec:2}, we define a multi-mode measurement device that includes the third beam splitter as marked in Fig.~\ref{fig3:DBSL}a. The resulting logic level is located just before the measurement device, where the generated 1D cluster state is coiled up, but not yet interfered with itself. A section of the cylindrical coiled up 1D cluster state at the logic level is shown in Fig.~\ref{fig3:DBSL}b. Here the horizontal direction follows the cluster state cylinder axis while the vertical direction corresponds to the circumference of the cylinder whose size is limited by the long delay line to $N$ temporal modes.

In the following we assume that the cylindrical cluster state has an even number of temporal modes in the circumference ($N$ is an even number which is necessary for the generated $\mathcal{H}$-graph state to be bipartite), each with a temporal mode index $k$. Every second temporal mode ($k+2i$ for $i\in\mathbb{Z}$) form \textit{wires} for computation along the cylinder (shaded area in Fig.~\ref{fig3:DBSL}b), while the remaining temporal modes ($k+2i-1$ for $i\in\mathbb{Z}$) are \textit{control modes} that are used to control couplings between wires. In this way we have $N/2$ wires, and thereby $N/2$ modes for computation. We will further assume that the number of wires is even, i.e. $N/2$ being an even number. As an example, the experimental realization of the DBSL in \cite{larsen19} had $N=12$ leading to 6 wires. Using an optical switch in the lower spatial mode at point $B$ in the logic level in Fig.~\ref{fig3:DBSL}a, an input state can be switched into the circuit. It corresponds to adding input states to the blue nodes in Fig.~\ref{fig3:DBSL}b. Optical switches have previously been demonstrated in quantum settings \cite{takeda19,larsen19b}.

By inducing certain phase rotations, $(-1)^i\theta_c$, of the control modes it is possible to create new edges along the wires as illustrated in Fig.~\ref{fig3:DBSL}c \cite{menicucci11a}. If these phase rotations are followed by measurement of the control modes in the $\x$-basis, the modes and their edges are ``deleted'', and we are consequently left with $N/2$ parallel wires suitable for single mode computation of $N/2$ modes. It is also worth noting that this combination of phase rotation and $\x$-measurement corresponds to measuring the quadrature $\x((-1)^i\theta_c)$ on each control mode individually: 
\begin{equation}\label{eq3:compBS}
	\includegraphics[width=0.89\linewidth]{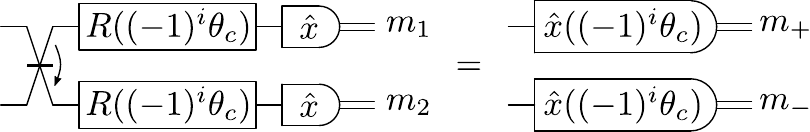},
\end{equation}
where the beam-splitter of the left-hand-side is the beam-splitter of the measurement device, and $m_\pm=(m_1\pm m_2)/\sqrt{2}$. As an example, we may consider the case of infinite squeezing as pictured in Fig.~\ref{fig3:DBSL}c. Here the edge weights of the wires tend towards the optimal values of $\pm1$ (where the sign alternates between neighbouring wires) by  choosing $\theta_c=\pi/4$. For finite, thus practical, squeezing levels, the induced wire edge weight is lower, while $\theta_c$ may be optimised for minimizing the gate noise. For simplicity, in the computation scheme presented here, we keep $\theta_c=\pi/4$ for all squeezing levels while in section \ref{sec3:variable_thetaC} we discuss the effect of varying $\theta_c$.

The projected wires in Fig.~\ref{fig3:DBSL}c are now suitable for single mode Gaussian computation: One computation step (one horizontal time step from temporal mode $k$ to $k+N$) corresponds to the generalized teleportation circuit in section \ref{sec2:gentele} with an input from the previous computation step, or switched into the cluster using an optical switch as previously mentioned. Similar to the generalized teleportation, the resulting operation of one single-mode computation step on a wire is
\begin{equation}\label{eq3:Singlemode}
	\Sq\left((-1)^i4t^2\right)\R\left(\frac{\theta_+}{2}\right)\Sq\left(\tan\frac{\theta_-}{2}\right)\R\left(\frac{\theta_+}{2}\right)\;,
\end{equation}
where $\theta_\pm=\theta_{Bk}\pm\theta_{Ak}$, and $t$ is the absolute edge weight in the logic dual-rail wire cluster state in Fig.~\ref{fig3:DBSL}b which equals $1/2$ in the infinite squeezing limit, while the case of finite squeezing is discussed in section~\ref{sec3:noise_analysis}. For derivation of (\ref{eq3:Singlemode}) see Appendix \ref{app:A}. The negative edge weight on every second wire (uneven $i$) leads to a $\pi$ phase rotation in each computation step which then cancels out in every second step, or can be compensated for in the required basis setting for the desired gate. As for the generalized teleportation, any Gaussian single-mode gate can be implemented in two steps.

\begin{figure}
	\includegraphics[width=0.9\linewidth]{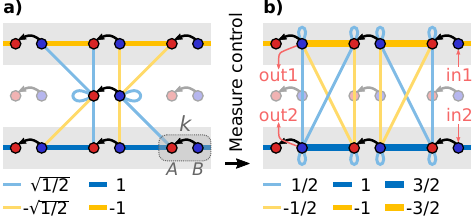}
	\caption{\label{fig3:CZ}(a) Logic cluster state after measuring all temporal control modes in Fig.~\ref{fig3:DBSL}(c) except control modes in temporal mode $k+N-1$. (b) Cluster state after further phase rotation and measurement of the remaining two control modes in (a), leading to direct edges between the neighbouring two wires. The edge weights shown here is for the case of infinite squeezing, $\theta_c=\pi/4$, and the central control modes further phase rotated by $\arctan(1/2)$, i.e. $g=1$ in Eq. (\ref{eq2:CZ}).}
\end{figure}

Now let us discuss how a two-mode gate can be implemented by coupling two neighboring wires. In Fig.~\ref{fig3:CZ}a all control modes except one has been measured in the basis $(-1)^i\theta_c=(-1)^i\pi/4$ in order to separate wires as described above---the remaining two central control modes in Fig.~\ref{fig3:CZ}a have only been phase shifted by $\pi/4$, but not measured. Phase rotating these remaining control modes further before measurements (i.e. measuring them in another bases than the neighbouring control modes), leads to coupling between the two neighbouring wires which is seen as direct edges in Fig.~\ref{fig3:CZ}b. In this way, controlling the measurement bases of a temporal control mode, together with the measurement bases of its neighbouring wires, a desired two-mode gates can be implemented. As an example, in the infinite squeezing limit of Fig.~\ref{fig3:CZ}, the base setting
\begin{equation}\label{eq2:CZ}
	\pmb{\theta}\equiv\begin{pmatrix}
		\theta_{Ak-2}\\ \theta_{Bk-2}\\ \theta_{Ak}\\ \theta_{Bk}\\ \theta_{Ak+N-2}\\ \theta_{Bk+N-2}\\ \theta_{Ak+N-1}\\ \theta_{Bk+N-1}\\ \theta_{Ak+N}\\ \theta_{Bk+N}
	\end{pmatrix}=
	\begin{pmatrix}
	(-1)^i 3\pi/8 \\
	-(-1)^{i}\pi/8 \\
	(-1)^i3\pi/8 \\
	-(-1)^i\pi/8 \\
	(-1)^i\pi/4-\arctan(g/2) \\
	-(-1)^i\pi/4 \\
	(-1)^i\pi/4+\arctan(g/2) \\
	(-1)^i\pi/4+\arctan(g/2) \\
	(-1)^i\pi/4-\arctan(g/2) \\
	-(-1)^i\pi/4
	\end{pmatrix}
\end{equation}
leads to an implementation of the gate $(\R(\pi/4)\otimes\R(\pi/4))\CZ(g)$ between the blue input modes in temporal mode $k-2$ and $k$, where $\R(\pi/4)\otimes\R(\pi/4)$ can be compensated for with the following single mode gates. 

In summary, we have now shown that a universal Gaussian gate set can be efficiently implemented on a DBSL cluster state: Single mode gates can be realized along parallel wires in the cluster while the two-mode controlled-Z gate can be realized between neighboring wires. 

\subsection{Gate noise analysis}\label{sec3:noise_analysis}
As mentioned previously, if the squeezed states used to construct the cluster state are infinitely squeezed, the gates will be realized perfectly without noise addition, thus without adding any processing errors. However, in a realistic setting, the degree of squeezing is finite which inevitably will result in processing noise. In the following we will be analysing the impact it has when using the DBSL for computation. 

Assuming that the two squeezed inputs states of the circuit in Fig.~\ref{fig3:DBSL}a have squeezed variances of $e^{-2r}$ in the $\x$- or $\p$-quadrature, the edge weights and self-loops of the coiled up 1D cluster state at the logic level becomes $\pm t=\pm\tanh(2r)/2$ and $i\varepsilon=i\,\sech(2r)$ respectively \cite{menicucci11a,larsen19}. Note that the existence of self-loops is a result of the finite input squeezing while $\sqrt{\varepsilon}$ can be considered as the effective momentum squeezing in the cluster state modes. 

The finite squeezing leads to two effects: Gate noise appearing in each computation step and distortion of the implemented gate. As seen in Eq.~(\ref{eq3:Singlemode}), for single mode gates the distortion is caused by an additional squeezing transformation, $\Sq((-1)^i\tanh^2(2r))$, on the output of each computation step. However, as for generalized teleportation, the unwanted squeezing transformation can be compensated for simply by tuning the basis settings. The gate noise (introduced in section \ref{sec2:gentele} and corresponding to the second term of Eq.~\eqref{eq2:trans}) of one single-mode computation step from temporal mode $k$ to $k+N$ (derived in Appendix \ref{app:A}) is represented by the following quadratures
\begin{equation*}
	\mathbf{N}(\p_{Ak-1},\p_{Ak},\p_{Ak+1},\p_{Bk+N-1},\p_{Bk+N},\p_{Bk+N+1})^T
\end{equation*}
with
\begin{equation*}
	\mathbf{N}=\begin{pmatrix}
\frac{1}{4t} & \frac{-1}{4t^2} & \frac{-1}{4t} & \frac{1}{4t} & 0 & \frac{1}{4t}\\
t & 0 & t & t & 1 & -t
\end{pmatrix},
\end{equation*}
leading to quadrature noise factors (introduced in section \ref{sec2:gentele}) of
\begin{equation}\begin{aligned}\label{eq3:INF}
	N_x&=\sum_jN_{1j}^2=\frac{1}{\tanh^4(2r)}+\frac{1}{\tanh^2(2r)}\\
	N_p&=\sum_jN_{2j}^2=\tanh^2(2r)+1
\end{aligned}\end{equation}
in $\x$ and $\p$ respectively. 

To avoid accumulating gate noise during computation, we consider the usage of GKP-encoded qubit states, in which the gate noise is translated into qubit errors by quadrature corrections after each implemented gate using auxiliary GKP-states as described in section \ref{sec2:errc}. To prevent erroneous computation, the qubits may then be error corrected by including a qubit error correction scheme in the computation. Within the GKP-encoded qubit subspace, a logic complete Clifford gate set is realized by the Gaussian gate set $\lbrace\I,\F,\Ps(1),\CZ(1)\rbrace$ on the bosonic modes. We therefore only consider the implementation of this gate set in the noisy cluster state. An additional non-Clifford gate in the GKP-qubit subspace completes the gate set for universal qubit computation, and is further discussed in section \ref{sec:6}.

Similar to the generalized teleportation circuit in section \ref{sec2:gentele} (Eq.~\eqref{eq2:gate}), but by substituting the edge weights, $t$, with $(-1)^i\tanh^2(2r)$, the single-mode $\I$-gate is implemented from temporal mode $k$ to $k+N$ with the basis setting
\begin{equation*}
	\begin{pmatrix}
		\theta_+\\\theta_-
	\end{pmatrix}_{I}=
	\begin{pmatrix}
		0\\
		(-1)^i2\arctan\left(\tanh^{-2}(2r)\right)
	\end{pmatrix}\;,
\end{equation*}
where $\theta_\pm=\theta_{Bk}\pm\theta_{Ak}$, and with gate noise variance of $N_x\varepsilon/2$ and $N_p\varepsilon/2$ in $\x$- and $\p$-quadrature respectively. The $\F$- and $\Ps(1)$-gate are implemented in two computation steps from mode $k$ to $k+2N$: Choosing
\begin{equation*}
	\begin{pmatrix}
		\theta_{+1}\\\theta_{-1}\\\theta_{+2}\\\theta_{-2}
	\end{pmatrix}_{F}
	=
	\begin{pmatrix}
		\pi/2\\
		\pi/2\\
		0\\
		2\arctan\left(\tanh^{-4}(2r)\right)
	\end{pmatrix}
\end{equation*}
implements $\F$ with equal gate noise variance of $(N_x+N_p)\varepsilon/2$ in $\x$ and $\p$, while 
\begin{equation*}
	\begin{pmatrix}
		\theta_{+1}\\\theta_{-1}\\\theta_{+2}\\\theta_{-2}
	\end{pmatrix}_{P}
	=
	\begin{pmatrix}
		\arctan 2\\
		-\arctan 2\\
		\pi/2\\
		\pi/2
	\end{pmatrix}
\end{equation*}
implements $\Ps(1)$ with gate noise variance of $2N_x\varepsilon/2$ and $2N_p\varepsilon/2$ in $\x$ and $\p$ respectively. Here $\theta_{\pm1}=\theta_{Bk}\pm\theta_{Ak}$ and $\theta_{\pm2}=\theta_{Bk+N}\pm\theta_{Ak+N}$. For the two-mode $\CZ$-gate, the gate distortion due to finite squeezing, and how it is compensated for, is less trivial. In the following, we search for basis settings that compensate for finite squeezing and optimizes the gate noise in order to minimize the error probability of the encoded qubit after quadrature corrections.

The GKP quadrature corrections can be realized by implementing the circuit in Eq.~(\ref{eq:ercCircuit}) at mode $B$ in the logic level in Fig.~\ref{fig3:DBSL} where the processed state is encoded. This may be challenging, as it requires tunable $\CZ(g)$-coupling strengths with $g=1$ when performing error correction, and otherwise $g=0$. An alternative is to occupy the free wires with ancillary $\ket{0_L}$ GKP-states, and then implement the required $\CZ(1)$-gates through measurements. However, with this approach, the error-correcting gate is subjected to the same kind of gate noise as we are trying to correct for in the encoded state. For simplicity, we assume successful implementation of the quadrature correction circuit in Eq.~\eqref{eq:ercCircuit} at the logic level using a supply of ancillary GKP-states with quadrature symmetric spike variance equal the variance of the resource squeezing, $\delta=e^{-2r}/2$.

\begin{figure*}
	\includegraphics[width=\linewidth]{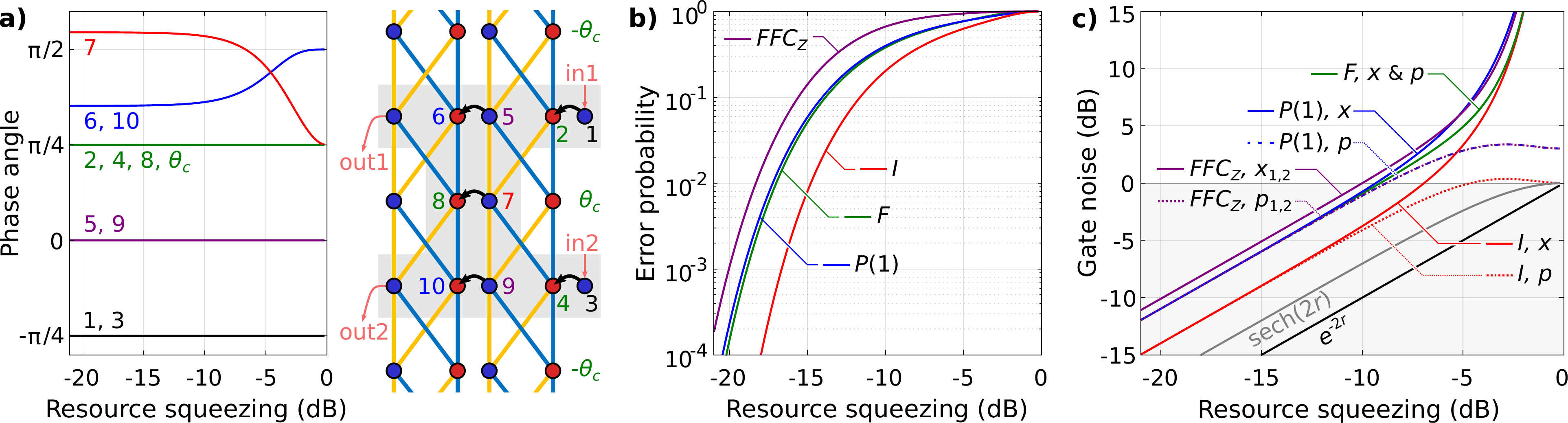}
	\caption{\label{fig3:DBSLnoise}(a) Basis settings as function of resource squeezing found by mimizing $f(\pmb{\theta})$ in Eq.~(\ref{eq3:objfun}) for implementing the $\CZ(1)$-gate as $(\F\otimes\F)\CZ(1)$ (in short $\F\F\CZ$) for even $i$ on the central control modes. The mode numbers are labelled on the graph to the right, where control modes outside the shaded area are measuring in basis $(-1)^i\theta_c=(-1)^i\pi/4$. For uneven $i$, $(\F\otimes\F^{-1})\CZ(1)$ is implemented with the same gate noise by changing the sign of the bases in mode 3, 4, 6, 7, 8, and 10. (b) Resulting error probabilities of Eq.~\eqref{eq2:Perr} for the $\I$- $\F$- and $\Ps(1)$-gate and the $(\F\otimes\F^{\pm1})\CZ(1)$-gate implemented with the basis settings in (a). (c) Gate noise responsible for the error probability in (b) together with appriximate ancillary GKP states of $e^{-2r}/2$ spike variance. Here the resource variance squeezing, $e^{-2r}$, and effective variance squeezing in the cluster state modes, $\sech(2r)$, is shown as well (black and grey respectively). The resource squeezing and gate noise in dB-scale is relative to vacuum variance of $1/2$. Note that the $\p$-quadrature gate noise of the $\Ps(1)$- and $\F\F\CZ$-gate overlap. For easy comparison, the gate noise here is shown in appendix \ref{app:C} together with gate noise of the schemes considered in section \ref{sec:4}.}
\end{figure*}

As discussed in section \ref{sec:3A} and illustrated in Fig.~\ref{fig3:CZ}, the $\CZ(1)$-gate between two wires is implemented in two computation steps, and while staying within the encoded qubit subspace, we are allowed to implement any $\CZ(1)$-gate with a bi-product of gates in $\lbrace\F,\Ps(1)\rbrace$ in order to minimize the resulting GKP-encoded qubit errors. The $\Ps(1)$-gate transforms quadratures as $(\x,\p)\rightarrow(\x,\x+\p)$, which, before adding gate noise, already leads to an increase of the spike variances in the GKP-encoded state as $(\delta,\delta)\rightarrow(\delta,2\delta)$, where the first and second index corresponds to variance in the $\x$- and $\p$-quadrature respectively. Thus adding $\Ps(1)$-gates to $\CZ(1)$ will hardly improve the error probability. On the other hand, the $\F$-gate transforms the quadratures as $(\x,\p)\rightarrow(\p,-\x)$, and the GKP-spike variance in each quadrature (before adding gate noise) is unchanged. Hence, we may improve the error probability if we can improve the resulting gate noise by adding $\F$-gates to the $\CZ(1)$-gate. We have investigated the gates $(\F^n\otimes\F^m)\CZ(1)$ for all $n,m\in\lbrace-1,0,1,2\rbrace$, and find that gates with $n,m=\pm1$ are optimal. We choose $n=1$ and $m=(-1)^i$, where the index $i$ denotes the control modes between the two coupled wires. The improvement on the $\CZ(1)$ gate noise by adding $\F\otimes\F^{\pm1}$ may be explained intuitively: The $\F$-gates rotate the states in computation during the two computation steps implementing $(\F\otimes\F^{\pm1})\CZ(1)$, and leads to the gate noise being better distributed on the quadratures, similar to the symmetrical distributed gate noise when implementing the single-mode $\F$-gate  as described above. The bi-product of $\F\otimes\F^{\pm1}$ can then be compensated for by applying the associated single mode gates after GKP error correction. In the following, we first consider the case for even $i$, and to shorten the notation we write $\F\F\CZ$ where the tensor product and $\CZ$-weight have been ignored. 

To implement the $\F\F\CZ$-gate between two neighbouring wires as in Fig.~\ref{fig3:CZ}, we adjust the basis setting $\pmb{\theta}$ in Eq.~(\ref{eq2:CZ}). Using a global search algorithm, we search for $\pmb{\theta}$ minimizing the objective function
\begin{equation}\label{eq3:objfun}
	f(\pmb{\theta})=||\mathbf{G}-\mathbf{T}||_1+w\log P_\text{err.}(\boldsymbol{\delta'},\delta)\;,
\end{equation}
where $\mathbf{G}$ and $\mathbf{T}$ are the symplectic matrices of the implemented gate, governed by $\pmb{\theta}$, and the target gate, $\F\F\CZ$ (see appendix \ref{app:A} for the procedure of calculating $\mathbf{G}$), $||\mathbf{A}||_1=\sum_{i,j}|A_{ij}|$ is the entrywise matrix 1-norm, and $P_\text{err.}$ is the error probability in Eq.~(\ref{eq2:Perr}). Here, $\delta=e^{-2r}/2$ for the ancillary GKP states, and $\boldsymbol{\delta'}=(2\delta,2\delta,\delta,\delta)^T+\boldsymbol{\sigma^2}$ for the $\F\F\CZ$-gate with gate noise variance $\boldsymbol{\sigma^2}=(N_{x1},N_{x2},N_{p1},N_{p2})^T\sech(2r)/2$ where $N_i$ are basis-dependent quadrature noise factors. The first term of $f(\pmb{\theta})$ in Eq.~(\ref{eq3:objfun}) is minimized for $\mathbf{G}=\mathbf{T}$, and thus helps us find the basis setting implementing the target gate \textbf{T}. Since multiple solutions, $\pmb{\theta}$, leading to $\textbf{G}=\textbf{T}$ may exist, we search for a solution that also minimizes the error probability, $P_\text{err.}$, which is the purpose of the second term in Eq.~\eqref{eq3:objfun}. To well resolve $P_\text{err.}$ close to $0$ we use the logarithm of $P_\text{err.}$, while the weight $w$ is varied in the range $10^{-8}$ to $1$ for different resource squeezing in order for the global search algorithm not to favour one term in (\ref{eq3:objfun}) while ignoring the other term. Finally, the objective function is considered successfully minimized only when the resulting gate is close to the target gate. To check this, we use the condition
\begin{equation*}
	||\mathbf{G}-\mathbf{T}||_1<10^{-5}\;,
\end{equation*}
with all results not satisfying this condition being discarded. Depending on the global search algorithm used, we are not guaranteed to find the best basis settings minimizing the error probability. However, repeating the algorithm many times with different $w$ and starting points increases the confidence of the resulting basis settings being optimal.

The resulting bases minimizing the objective function $f(\pmb{\theta})$ for the $\F\F\CZ$ target gate (with even $i$ for the central control mode) is presented in Fig.~\ref{fig3:DBSLnoise}a for different resource squeezing levels as input in Fig.~\ref{fig3:DBSL}a. In the following we will refer to the mode numbering labelled in Fig.~\ref{fig3:DBSLnoise}a. According to Eq.~(\ref{eq3:Singlemode}), with $(\theta_{1,3},\theta_{2,4})=(-\pi/4,\pi/4)$ in the first computation step, ignoring the coupling between wires, the input mode is simply teleported to the second computation step with a bi-product (beside displacement) of $\Sq(\tanh^2(2r))$. Here, control mode 8 is measured in the same $\theta_c=\pi/4$ basis used for separating wires, while control mode 7 is measured in a different basis in order to couple the two wires. With the combined basis setting of mode 5, 6, 7, 9 and 10 the bi-product squeezing of the first step is compensated, and the $\F\F\CZ$-gate is implemented. Finally, for uneven $i$ on the control modes between the two coupled wires, the $(\F\otimes\F^{-1})\CZ(1)$-gate is implemented by changing the sign on mode 3, 4, 6, 7, 8, and 10. The resulting gate noise and error probability is the same as for $\F\F\CZ$ with even $i$.

After quadrature correction in the GKP-scheme the resulting error probability of the above described basis settings for the $\I$, $\F$ and $\Ps(1)$ single-mode gates and the two-mode $\F\F\CZ$-gate is shown in Fig.~\ref{fig3:DBSLnoise}b. As expected, the error probability is seen to go towards 0 for increasing resource squeezing, and towards 1 for vanishing squeezing. Furthermore, the $\F\F\CZ$-gate is seen to have the highest error probability due to four successful quadrature corrections necessary to avoid qubit error, while the $\I$-gate leads to the lowest error probability as it is implemented in a single computation step. In section \ref{sec5} these error probabilities are compared with error probabilities when using other relevant cluster states and computing schemes.

To gain a better understanding of the error probabilities, we consider the responsible gate noise. The gate noise variance, for the basis settings in Fig.~\ref{fig3:DBSLnoise}a and described above, is plotted in Fig.~\ref{fig3:DBSLnoise}c. In the large squeezing limit, the effective variance squeezing in the cluster state modes of $\sech(2r)$ is a factor of 2 ($\SI{3}{dB}$) larger than the resource variance squeezing of $e^{-2r}$, which is the cost of preparing the cluster state with off-line squeezing \cite{gu09}. The $\I$-gate, implemented in a single computation step, has a gate noise in the range two times higher than the effective squeezing due to $N_x,N_p\rightarrow2$ for $r\rightarrow\infty$. The $\F$- and $\Ps(1)$-gate have further gate noise of around a factor two, since they are implemented in two computation steps. Finally, the $\F\F\CZ$-gate, also implemented in two computation steps, has similar gate noise, but slightly higher due to the noise of an additional control mode included in the gate to couple two neighbouring wires. The gate noise is in general asymmetric in the quadratures (besides for the $\F$-gate with equal noise factor in the two quadratures), also for the $\F\F\CZ$-gate with optimized basis settings: Since $P_\text{err.}$ in Eq.~(\ref{eq2:Perr}) rely on the product of quadrature correction success, and because the encoded GKP spike noise is also asymmetric after the $\F\F\CZ$- and $\Ps(1)$-gate, the error probability is not necessary minimum with quadrature symmetric gate noise, and at low squeezing we see the majority of the gate noise in one quadrature. Finally, in the vanishing squeezing limit the gate noise diverges. To understand this, consider the Wigner function transformation of the generalized teleportation in Eq.~\eqref{eq2:Wigner}: With the diverging gate noise variance, the Wigner function is convoluted with infinitely broad Gaussian functions in $\x$ and subjected to corresponding delta function envelopes in $\p$, erasing all information of the encoded state. Together with convolutions in the $\p$-quadrature and corresponding envelopes in the $\x$-quadrature, the Wigner function is ensured to go towards vacuum for $\SI{0}{dB}$ resource squeezing. This is further described in appendix \ref{app:B} with the Wigner function transformation of single-mode gates on the DBSL.

\subsection{Variable control mode basis}\label{sec3:variable_thetaC}
For simplicity, so far we fixed control mode basis to $\theta_c=\pi/4$, which only leads to unity wire edge weight in the infinite squeezing limit. Allowing variable $\theta_c$ implements
\begin{equation*}
	\Sq\left((-1)^i4t^2\tan\theta_c\right)\R\left(\frac{\theta_+}{2}\right)\Sq\left(\tan\frac{\theta_-}{2}\right)\R\left(\frac{\theta_+}{2}\right)
\end{equation*}
single-mode gates in each computation step with
\begin{equation*}
	\mathbf{N}=\begin{pmatrix}
	\frac{1}{4t} & \frac{-1}{4t^2\tan\theta_c} & \frac{-1}{4t} & \frac{1}{4t} & 0 & \frac{1}{4t}\\
	t\tan\theta_c & 0 & t\tan\theta_c & t\tan\theta_c & 1 & -t\tan\theta_c
	\end{pmatrix}
\end{equation*}
for the gate noise leading to
\begin{equation*}\begin{aligned}
	N_x&=\sum_jN_{1j}^2=\frac{1}{\tanh^4(2r)\tan^2\theta_c}+\frac{1}{\tanh^2(2r)}\\
	N_p&=\sum_jN_{2j}^2=\tanh^2(2r)\tan^2\theta_c+1
\end{aligned}\end{equation*}
noise factors. As a result, by varying $\theta_c$ we are able to distribute the gate noise between the quadratures in order to minimize the GKP-encoded qubit errors.

\begin{figure}
	\includegraphics[width=\linewidth]{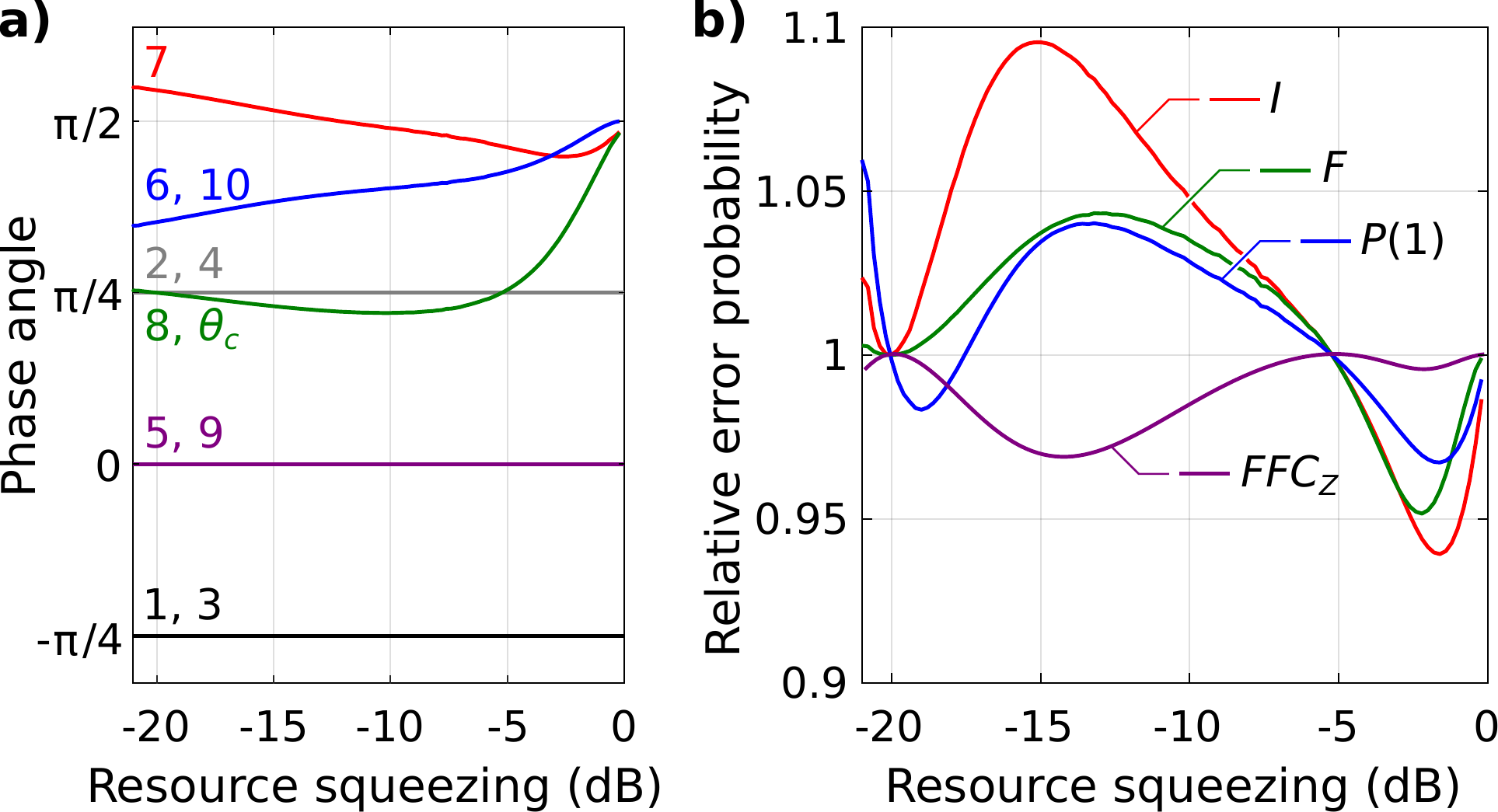}
	\caption{\label{fig3:variable_thetaC}(a) Basis setting implementing $(\F\otimes\F)\CZ(1)$ (shortened $\F\F\CZ$) for even $i$ with variable control basis $\theta_c$ minimizing the error probability of Eq.~\eqref{eq2:Perr}, while $(\F\otimes\F^{-1})\CZ(1)$ is implemented for uneven $i$ by changing the sign on mode 3, 4, 6, 7, 8 and 10. Here the mode numbering used is that of Fig.~\ref{fig3:DBSLnoise}a. (b) Resulting error probabilities using variable $\theta_c$ in (a) optimized for $\F\F\CZ$, relative to the corresponding error probabilities in Fig.~\ref{fig3:DBSLnoise}b for fixed $\theta_c=\pi/4$.}
\end{figure}

To prevent unwanted couplings between wires, $\theta_c$ needs to be the same for all gates. With the two-mode $\CZ(1)$-gate being the gate of largest error probability, we may optimize $\theta_c$ to minimize the error probability of  the $(\F\otimes\F^{\pm1})\CZ(1)$-gate. In Fig.~\ref{fig3:variable_thetaC} optimized basis settings, and corresponding error probabilities relative to the error probabilities for fixed $\theta_c=\pi/4$, is shown as function of resource squeezing when including $\theta_c$ in the objective function in Eq.~\eqref{eq3:objfun}. The error probability for the $(\F\otimes\F^{\pm1})\CZ(1)$-gate, for which $\theta_c$ is optimized, is seen at best to decrease to $0.97$ of the error probability with fixed $\theta_c$, and thus the gain of variable $\theta_c$ is little. Furthermore, since $\theta_c$ is only optimized for the $(\F\otimes\F^{\pm1})\CZ(1)$-gate, for some ranges of resource squeezing, the error probabilities for the $\I$-,  $\F$- and $\Ps(1)$-gate is seen to becomes worse. In conclusion, there may be a small advantage of optimizing $\theta_c$, but this depends on the amount of resource squeezing available, and what gates dominate the quantum algorithm to be implemented.

\section{Other cluster states}\label{sec:4}
Besides the DBSL, there are three other interesting cluster states with corresponding self-inverse and bipartite $\mathcal{H}$-graph states and thus realizable with off-line squeezing and linear optics. It counts the quad-rail lattice (QRL) \cite{menicucci11b} with the efficient computation scheme in Ref. \cite{alexander16a}; the bilayer square lattice (BSL) \cite{alexander16b,alexander18}, also with an efficient computation scheme; and the recently demonstrated cluster state by Asavanant \textit{et al.} \cite{asavanant19}. In the following, we refer to this last cluster state as the modified bilayer square lattice (MBSL) since computation on this state is similar to computation on the BSL with few modifications. Below, we summarize the computation schemes for each cluster state focusing on the $\lbrace\I,\F,\Ps(1),\CZ(1)\rbrace$ gate set which, together with $\sqrt{\pi}$ displacements in $\x$- and $\p$-quadratures, constitute a universal Clifford gate set in the GKP-encoded qubit subspace. Here, we apply the same search for basis settings that optimize the gate noise in order to minimize qubits errors---as figure of merit, we use the the error probability of Eq.~\eqref{eq2:Perr}. For easy comparison, the figures summarizing the different considered schemes and the resulting gate noise are also put together in appendix \ref{app:C}. The resulting error probabilities are then compared with the error probabilities for the DBSL in the section \ref{sec5}, while universality through the implementation of a non-Clifford gate in the various schemes is discussed in section \ref{sec:6}.

\subsection{Bilayer square lattice}
\begin{figure*}
	\includegraphics[width=\linewidth]{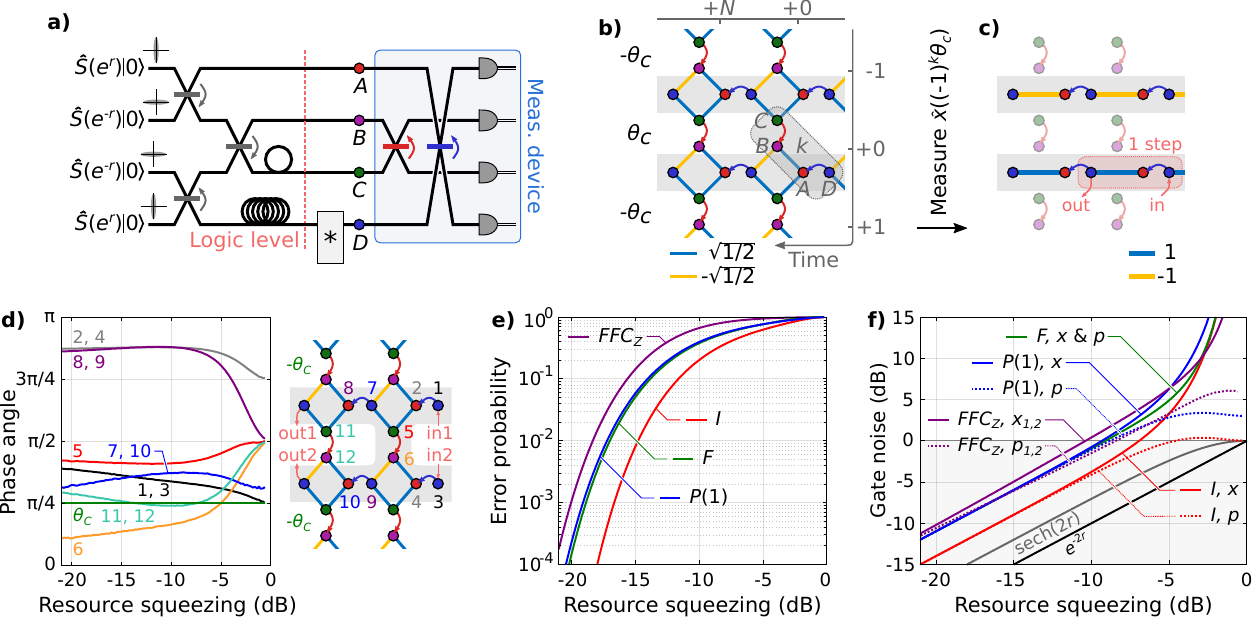}
	\caption{\label{fig4:BSL} Bilayer square lattice (BSL): (a) Experimental setup for generating the $\mathcal{H}$-graph state corresponding to the BSL cluster state \cite{alexander16b,alexander18}. Here the device marked by asterisk is described in Fig.~\ref{fig3:DBSL} for the DBSL, and represents a switch for switching in and out states or a GKP quadrature correction circuit. The logic level in which the computation takes place is marked, and the corresponding logic cluster state is shown in (b) with arrows representing the beam splitters of the measurement device, while in (c) the logic cluster state is projected into wires for computation. The edge weights shown here are in the limit of infinite squeezing and $\theta_c=\pi/4$. (d) Basis setting implementing the $(\F\otimes\F^{-1})\CZ(1)$-gate for even temporal modes $k$ with a minimum error probability. For uneven $k$, $(\F\otimes\F)\CZ(1)$ is implemented by changing the sign of the bases. The error probability in Eq.~\eqref{eq2:Perr} of the single mode $\I$-, $\F$-, $\Ps(1)$-gate with $\theta_c=\pi/4$, and the $(\F\otimes\F^{\pm1})\CZ(1)$-gate, are presented in (e) with the corresponding gate noise shown in (f)---here, $\F\F\CZ$ is short for $(\F\otimes\F^{\pm1})\CZ(1)$. The experimental setup, logic cluster state and its projection into wires, and the resulting gate noise, is shown together with the other considered schemes in appendix \ref{app:C} for easy comparison.}
\end{figure*}

The 2-dimensional BSL can be generated in the time-frequency domain using a single optical parametric oscillator \citep{alexander16b} or solely in time domain using four squeezing sources \cite{alexander18} as summarized in Fig.~\ref{fig4:BSL}a. We emphasize that the time-only encoding of the BSL in \citep{alexander18} is not necessarily more favourable than the frequency-time encoding in \citep{alexander16b}---one may even argue that the frequency-time encoding has a better scaling performance. Here, we simply present the time-only encoded version of the setup, since it is comparable to that of the QRL and the MBSL, but it is important to note that the analysis presented in this work holds also for the time-frequency encoded BSL.

The setup in Fig.~\ref{fig4:BSL}a produces a self-inverse and bipartite $\mathcal{H}$-graph state, which under phase rotations is transformed into a cluster state. An efficient universal computation scheme is well described by Alexander \textit{et al.} \cite{alexander16b,alexander18} in the language of macronodes in which each macronode corresponds to the logic level marked in Fig.~\ref{fig4:BSL}a. The computation takes place at this level and the logic cluster state consists of square cluster states as presented in Fig.~\ref{fig4:BSL}b with $\pm t=\pm\tanh(2r)/\sqrt{2}$ edge weight and $i\varepsilon=i\sech(2r)$ self-loops. The measuring system comprises two joint measurements for each temporal mode $k$: A joint measurement of the control modes $B$ and $C$ in basis $(-1)^k\theta_c$ to project the cluster state into wires as shown in Fig.~\ref{fig4:BSL}c, and a joint measurement of the wire modes $A$ and $D$ to implement gates on these wires. As for the DBSL, we find that $\theta_c=\pi/4$ is near optimal. Measuring wire modes $A$ and $B$ of temporal mode $k$ in bases $\theta_{Ak}$ and $\theta_{Dk}$ implements the single mode gate 
\begin{equation}\label{eq4:BSLsinglemode}
	\Sq\left((-1)^{k+1}2t^2\right)\R\left(\frac{\theta_+}{2}\right)\Sq\left(\tan\frac{\theta_-}{2}\right)\R\left(\frac{\theta_+}{2}\right)
\end{equation}
from temporal mode $k$ to $k+N$ (1 computation step) where $\theta_\pm=\theta_{Dk}\pm\theta_{Ak}$. The resulting gate noise for one computation step is $\textbf{N}(\p_{Ak},\p_{Bk},\p_{Ck+1},\p_{Dk+N})^T$ where
\begin{equation}\label{eq4:BSL_N}
	\textbf{N}=\begin{pmatrix}
		\frac{1}{2t^2} & \frac{1}{2t} & -\frac{1}{2t} & 0\\
		0 & -t &-t & 1
	\end{pmatrix}\;.
\end{equation}
Thus, the variance of the gate noise added to the output quadratures in each computation step is $N_x\varepsilon/2$ and $N_p\varepsilon/2$ for the $\x$- and $\p$-quadrature respectively, where
\begin{equation*}\begin{aligned}
	N_x&=\frac{1}{\tanh^4(2r)}+\frac{1}{\tanh^2(2r)}\\
	N_p&=\tanh^2(2r)+1
\end{aligned}\end{equation*}
are quadrature noise factors, $\sum_jN_{ij}^2$, introduced in section \ref{sec2:gate_noise}, and we note that they are identical to the noise factors of the DBSL. The $\I$-gate is implemented in a single computation step by choosing
\begin{equation*}
	\begin{pmatrix}
		\theta_+\\ \theta_-
	\end{pmatrix}_I=
	\begin{pmatrix}
		0\\ (-1)^{k+1}2\arctan\left(\tanh^{-2}(2r)\right)
	\end{pmatrix}\;.
\end{equation*}
The $\F$- and $\Ps(1)$-gate are implemented in two computation steps from temporal mode $k$ to $k+2N$. Choosing basis settings
\begin{equation*}
	\begin{pmatrix}
	\theta_{+1}\\ \theta_{-1}\\ \theta_{+2}\\ \theta_{-2}
	\end{pmatrix}_F=
	\begin{pmatrix}
	\pi/2 \\ \pi/2 \\ 0 \\ 2\arctan\left(\tanh^{-4}(2r)\right)
	\end{pmatrix}
\end{equation*}
$\F$ is implemented with equal gate noise variance of $(N_x+N_p)\varepsilon/2$ in $\x$- and $\p$-quadrature, while $\Ps(1)$ is realized with 
\begin{equation*}
	\begin{pmatrix}
	\theta_{+1}\\ \theta_{-1}\\ \theta_{+2}\\ \theta_{-2}
	\end{pmatrix}_P=
	\begin{pmatrix}
	\arctan 2 \\ -\arctan 2 \\ \pi/2 \\ \pi/2
	\end{pmatrix}
\end{equation*}
resulting in gate noise variances of $2N_x\varepsilon/2$ and $2N_p\varepsilon/2$ in $\x$- and $\p$-quadrature, respectively. Here $\theta_{\pm1}=\theta_{Dk}\pm\theta_{Ak}$ and $\theta_{\pm2}=\theta_{Dk+N}\pm\theta_{Ak+N}$. Notice the similarity with the DBSL: The basis settings and gate noises are identical, and the BSL and DBSL are expected to perform single mode gates equally well.

Measuring control modes $B$ and $C$ of one temporal mode in different bases leads to coupling between the two neighbouring wires and thus allow for the implementation of two-mode gates. In ref. \cite{alexander16b}, the basis setting for implementing the $\CZ(g)$-gate is given for the case of infinite squeezing. Here, we extend this analysis by 
searching the basis setting that minimizes the error probability of two encoded qubits after the $\CZ(1)$-gate for the more relevant case of finite squeezing. To do so, we use the same technique as for the DBSL by minimizing the objective function in Eq.~\eqref{eq3:objfun}. Note that to compensate for finite squeezing distortion (as $\Sq(\pm\tanh^2(2r))$ in Eq.~\eqref{eq4:BSLsinglemode} for single mode gates), two computation steps are required to implement $\CZ(1)$. For all $(\F^n\otimes\F^m)\CZ(1)$-gates with $n,m=0,1,2,3$ we find the lowest error probability for $n,m=\pm1$ and we choose $(n,m)=(1,(-1)^{k+1})$ where $k$ is the temporal mode index of the control modes coupling the two wires. The resulting basis settings implementing the $(\F\otimes\F^{-1})\CZ(1)$-gate are shown in Fig.~\ref{fig4:BSL}d for even $k$, while for uneven $k$ the $(\F\otimes\F)\CZ(1)$-gate is implemented with equal error probability by changing the sign of all bases in Fig.~\ref{fig4:BSL}d. 
In case we allow for a variable $\theta_c$ in the objective function of Eq.~\eqref{eq3:objfun}, we find no improvement of the error probability, and we conclude there will be no gain of a variable $\theta_c$ when implementing the $(\F\otimes\F^{\pm1})\CZ(1)$-gate.

Note how the basis settings in Fig.~\ref{fig4:BSL}d, different from the DBSL in Fig.~\ref{fig3:DBSLnoise}a and the MBSL later in Fig.~\ref{fig4:Asavanant}d, seem to depend on the resource squeezing in the full shown squeezing range. The reason for this is that there exist multiple solutions for basis settings that implement a desired $\CZ(1)$-gate with a minimum error probability. The same is the case for the DBSL and MBSL, however, in Fig 4d and 7d a more consistent solution set of basis settings as function of resource squeezing is shown, while here for the BSL a slightly inconsistent solution set is shown. This effect often occurs when unnecessarily large degrees of freedom in the basis settings are used when minimizing the objective function in Eq.~\eqref{eq3:objfun}. However, this does not mean that the basis settings in Fig.~\ref{fig4:BSL}d are not optimal, but is rather an example of the existence of multiple basis setting solutions and proper use of Eq.~\eqref{eq3:objfun} to derive a suitable solution for a given experimental implementation.

The resulting error probabilities of Eq.~\eqref{eq2:Perr} when correcting the quadratures after the $\I$-, $\F$-, $\Ps(1)$- and $(\F\otimes\F^{\pm1})\CZ(1)$-gate as described above are presented in Fig.~\ref{fig4:BSL}e. As expected, the two-mode $\CZ(1)$-gate is seen to have the highest error probability since four successful quadrature corrections are necessary to avoid inducing an error on the encoded qubits. Finally, the gate noise variances are presented in Fig.~\ref{fig4:BSL}f, and here we clearly see similar behavior as for the DBSL: For infinite squeezing, the $\I$-gate in one computation step has a gate noise variance of twice the effective variance squeezing, $\sech(2r$) (as $N_x,N_p\rightarrow2$ when $r\rightarrow\infty$), while the $\F$- and $\Ps(1)$-gates implemented in two computation steps have gate noise variances of four times that. In the other extreme of vanishing squeezing, the gate noise diverges in the $\x$-quadrature, thereby erasing all information of the encoded state as previously explained for the DBSL. This can also be seen from the corresponding Wigner function transformation in appendix \ref{app:B}.

\subsection{Modified bilayer square lattice}
\begin{figure*}
	\includegraphics[width=\linewidth]{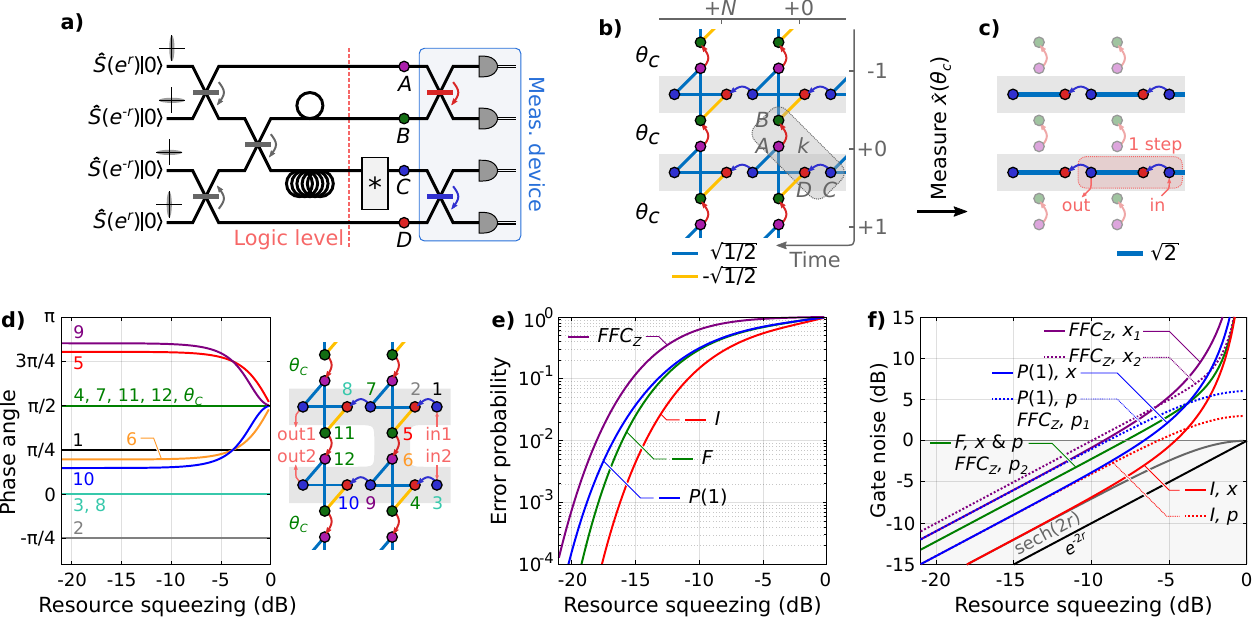}
	\caption{\label{fig4:Asavanant}Modified bilayer square lattice (MBSL): (a) Experimental setup for generating the $\mathcal{H}$-graph state corresponding to the MBSL cluster state by Asavanant \textit{et al.} \cite{asavanant19}. Here the device marked by asterisk is described in Fig.~\ref{fig3:DBSL} for the DBSL, and represents a switch for switching in and out states or a GKP quadrature corrections circuit. The logic level in which the computation takes place is marked, and the corresponding logic cluster state is shown in (b) with arrows representing the beam splitters of the measurement device, while in (c) the logic cluster state is projected into wires for computation. The edge weights shown here are in the limit of infinite squeezing and $\theta_c=\pi/2$. (d) shows a basis setting implementing the $(\F\otimes\F)\CZ(1)$-gate with a minimum error probability. The error probability in Eq.~\eqref{eq2:Perr} of the single-mode $\I$-, $\F$-, and $\Ps(1)$-gate with $\theta_c=\pi/2$, and the $(\F\otimes\F^{\pm1})\CZ(1)$-gate, is presented in (e) with the corresponding gate noise shown in (f)---here $\F\F\CZ$ is short for $(\F\otimes\F)\CZ(1)$. Note that the gate noise variance in the $\p$-quadratures of each mode for the $(\F\otimes\F)\CZ(1)$-gate each equals that of the $\Ps(1)$ and $\F$ gate. The experimental setup, logic cluster state and its projection into wires, and the resulting gate noise, is shown together with the other considered schemes in appendix \ref{app:C} for easy comparison.}
\end{figure*}

The experimental setup of the MBSL cluster state, recently generated by Asavanant \textit{et al.} \cite{asavanant19} and summarized in Fig.~\ref{fig4:Asavanant}a, is very similar to the setup of the all-time encoded BSL in Fig.~\ref{fig4:BSL}a, and we can therefore adopt the computation scheme for the BSL with only a few changes. The corresponding cluster state at the logic level is shown in Fig.~\ref{fig4:Asavanant}b in which we see that the square clusters of the BSL have been replaced with ``butterfly'' clusters. As for the BSL, the edge weight and self-loops are $\pm t=\pm\tanh(2r)/\sqrt{2}$ and $i\varepsilon=i\sech(2r)$ respectively. The spatial modes $C$ and $D$ of each temporal mode $k$ constitutes wire modes, while $A$ and $B$ are control modes. In contrast to the square clusters in the BSL, the butterfly clusters already contain direct edges in the wires before potential phase rotation of the control modes. Thus we can directly ``delete'' the control modes by measuring them in the $\x$-basis, i.e. $\theta_c=0$, and implement the operations
\begin{equation*}
	\Sq(t)\R\left(\frac{\theta_+}{2}\right)\Sq\left(\tan\frac{\theta_-}{2}\right)\R\left(\frac{\theta_+}{2}\right)\quad(\theta_c=0)
\end{equation*}
in one computation step from temporal mode $k$ to $k+N$ with $\theta_\pm=\theta_{Ck}\pm\theta_{Dk}$. The resulting gate noise is $\textbf{N}(\p_{Dk},\p_{Ak},\p_{Bk+1},\p_{Ck+N})^T$ with
\begin{equation*}
	\textbf{N}=\begin{pmatrix}
		-\frac{1}{t} & 0 & 0 & 0\\ 0&0&0&1
	\end{pmatrix}\quad(\theta_c=0)
\end{equation*}
such that the gate noise variance is $N_x\varepsilon/2$ and $N_p\varepsilon/2$ in $\x$- and $\p$-quadrature respectively with quadrature noise factors of
\begin{equation*}
	\begin{matrix}
	N_x=2/\tanh^2(2r)\vspace*{2mm}\\
	N_p=1\hspace*{1.65cm}	
	\end{matrix}\quad(\theta_c=0)\;.
\end{equation*}

Alternatively, we can measure the control modes in the $\p$-basis, i.e. $\theta_c=\pi/2$, rearranging the edge weights of the butterfly cluster states to increase the edge weight between wire modes as shown in Fig.~\ref{fig4:Asavanant}c. In this case, the operation
\begin{equation*}
	\Sq(2t)\R\left(\frac{\theta_+}{2}\right)\Sq\left(\tan\frac{\theta_-}{2}\right)\R\left(\frac{\theta_+}{2}\right)\quad(\theta_c=\pi/2)
\end{equation*}
is implemented with the gate noise,
\begin{equation}\label{eq4:A_N}
	\textbf{N}=\begin{pmatrix}
		-\frac{1}{2t} & -\frac{1}{2t} & 0 & 0\\ 0&0&-1&1
	\end{pmatrix}\quad(\theta_c=\pi/2)
\end{equation}
such that
\begin{equation*}
	\begin{matrix}
	N_x=1/\tanh^2(2r)\vspace*{2mm}\\
	N_p=2\hspace*{1.63cm}	
	\end{matrix}\quad(\theta_c=\pi/2)\;.
\end{equation*}
Other values of $\theta_c$ are also possible, but in this case the implemented gate as well as gate noise is less trivial. However, in the later analysis of the $\CZ(1)$-gate we do find that $\theta_c=\pi/2$ is indeed optimal. Notice that---unlike the BSL with square cluster states---all control modes are measured in the same basis without an alternating sign for different temporal modes. This is because the wire modes are directly connected with equal edge weights for all temporal modes, or connected with three edges through two control modes, while for the square cluster states, wire modes have two connections, each through a single control mode, but with different sign on the edge weights depending on whether the control mode is in the next or previous temporal mode.

For $\theta_c=\pi/2$, the basis setting
\begin{equation*}
	\begin{pmatrix}
		\theta_+\\ \theta_-
	\end{pmatrix}_I=
	\begin{pmatrix}
	0\\2\arctan\left(\tanh^{-1}(2r)/\sqrt{2}\right)
	\end{pmatrix}
\end{equation*}
with $\theta_\pm=\theta_{Ck}\pm\theta_{Dk}$ implements the $\I$-gate in one computation step from temporal mode $k$ to $k+N$ and gate noise variance $N_x\varepsilon/2$ and $N_p\varepsilon/2$ in $\x$- and $\p$-quadrature respectively. The basis setting
\begin{equation*}
	\begin{pmatrix}
	\theta_{+1}\\\theta_{-1}\\\theta_{+2}\\\theta_{-2}
	\end{pmatrix}_F=\begin{pmatrix}
	\pi/2\\\pi/2\\0\\2\arctan\left(\tanh^{-2}(2r)/2\right)
	\end{pmatrix}
\end{equation*}
implements the $\F$-gate in two computation steps with equal $(N_x+N_p)\varepsilon/2$ gate noise variance in $\x$- and $\p$-quadrature, while
\begin{equation*}
	\begin{pmatrix}
	\theta_{+1}\\\theta_{-1}\\\theta_{+2}\\\theta_{-2}
	\end{pmatrix}_P=\begin{pmatrix}
	\arctan2\\-\arctan2\\\pi/2\\\pi/2
	\end{pmatrix}
\end{equation*}
implements the $\Ps(1)$-gate in two computation steps with $2N_x\varepsilon/2$ and $2N_p\varepsilon/2$ gate noise variance in $\x$- and $\p$-quadrature respectively. Here, $\theta_{\pm1}=\theta_{Ck}\pm\theta_{Dk}$ and $\theta_{\pm2}=\theta_{Ck+N}\pm\theta_{Dk+N}$ when implementing $\F$ and $\Ps(1)$ from temporal mode $k$ to $k+2N$.

To couple pairs of wires for the implementation of a two-mode gate, one measures the control modes $A$ and $B$ of one temporal mode $k$ in different bases by which a coupling between the wires in temporal modes $k-1$ and $k$ is induced. The $\CZ(1)$-gate is again implemented in two computation steps, and similar as for the BSL and the DBSL, we search the basis setting that minimizes the objective function in Eq.~\eqref{eq3:objfun} and thus the error probability in Eq.~\eqref{eq2:Perr} of that particular gate. Again, we need to investigate all $(\F^n\otimes\F^m)\CZ(1)$-gates for $n,m=0,1,2,3$ and find $n=m=1$ to be optimal. The resulting basis setting is shown in Fig.~\ref{fig4:Asavanant}d where $\theta_c=\pi/2$ is found to be optimal. Note that, unlike the DBSL and BSL, this basis setting is independent on the temporal mode index $k$, as the control basis does not have an alternating sign governed by $k$.

The resulting error probability of the single mode $\I$-, $\F$- and $\Ps(1)$-gate, and the two-mode $(\F\otimes\F)\CZ(1)$-gate, with the basis settings described above and in Fig.~\ref{fig4:Asavanant}d, is shown in Fig.~\ref{fig4:Asavanant}e. The single mode gates are all seen to have a lower error probability than in computations with the DBSL and BSL cluster states. This is explained by the lower quadrature noise factors, $N_x$ and $N_p$, due to the structure of the butterfly cluster states with initial edges between wire modes before projecting the logic cluster state into wires. As expected, due to the four quadrature corrections, the error probability of the $(\F\otimes\F)\CZ(1)$-gate is largest. Gate noise variances are shown in Fig.~\ref{fig4:Asavanant}f. For single mode gates, in general we see lower gate noise variance than for the DBSL and BSL, and in the large squeezing limit where $N_x\rightarrow1$ for $r\rightarrow\infty$ and $\theta_c=\pi/2$, while $N_p=2$, we see the gate noise variances in $\x$-quadratures of the $\I$-gate to equal the effective squeezing variance of $\sech(2r)$. For vanishing squeezing, the gate noise variance diverges in the $\x$-quadrature, erasing all information of the encoded state as is also the case for computing with BSL and DBSL (also eluded by the Wigner function transformation in appendix \ref{app:B}). Notice that, unlike the DBSL and BSL, the gate noise of the $(\F\otimes\F)\CZ(1)$-gate is not symmetric in the quadratures of the two modes.

\subsection{Quad-rail lattice}
\begin{figure*}
	\includegraphics[width=\linewidth]{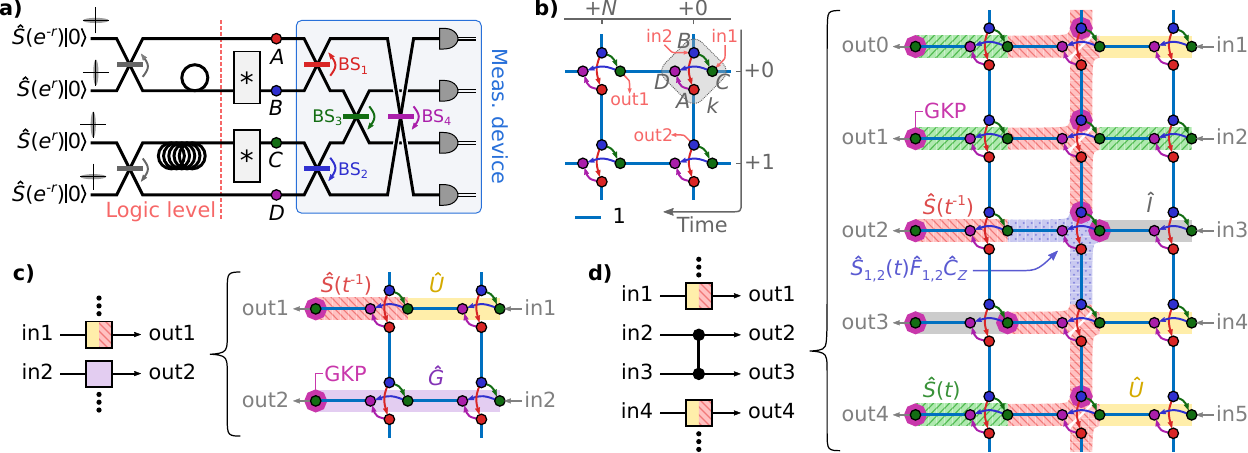}
	\caption{\label{fig4:QRL}Quad-rail lattice (QRL): (a) Experimental setup for generating the $\mathcal{H}$-graph state corresponding the the QRL cluster state. Here, the devices marked by asterisk are described in Fig.~\ref{fig3:DBSL} and represent optical switches switching in and out states or GKP quadrature correction circuits as in Eq.~\eqref{eq:ercCircuit}. The logic level in which the computation takes place is marked, with the corresponding logic cluster state shown in (b) with arrows representing the beam-splitter network of the measurement device. The edge weight shown is in the infinite squeezing limit, while for finite squeezing it is $t=\tanh(2r)$, while $i\varepsilon=i\sech(2r)$ self-loops are present on all nodes. Temporal mode indices are written in grey, while a single computation step is marked with input modes `in1' and `in2' and output modes `out1' and `out2'. (c) Example of single mode computation along the cluster state cylinder by restricting the bases with $\theta_{Ak}=\theta_{Dk}$ and $\theta_{Bk}=\theta_{Ck}$, thereby implementing $\hat{U}$ in Eq.~\eqref{eq4:U} in each computation step. With two computation steps, the $\Sq(t)$ distortion in Eq.~\eqref{eq4:U} due to finite squeezing can be compensated in the second step by implementing $\Sq(t^{-1})$. Or, more generally, any single-mode Gaussian gate can be implemented as $\hat{G}=\hat{U}_1\hat{U}_2$. Here, input modes in spatial modes $B$ (blue) are ignored. After implemention of gates, the output modes marked with a pink circle are quadrature corrected. (d) Example of implementing the $(\F\otimes\F)\CZ(1)$-gate between input modes `in2' and `in3'. Since all gates are implemented on pairs of modes, first one computation step is required to guide the in2 and in3 modes to the $(\F\otimes\F)CZ(1)$-gate, while gates of the form $\Sq(t^{-1})\hat{U}$ can be implemented on other computation modes. After the $(\F\otimes\F)CZ(1)$-gate, all computation modes are aligned to the same vertical position in the lattice using $\I$-gates (notice that $\Sq(t^{\pm1})\Sq(t^{\mp1})=\I$). To prevent accumulating gate noise, GKP quadrature correction are performed after every implemented gate on modes marked with pink circle. The experimental setup and logic cluster state is shown together with the other considered schemes in appendix \ref{app:C} for easy comparison.}
\end{figure*}

In ref. \cite{menicucci08}, it was proposed to generate a cluster state with a quad-rail lattice (QRL) structure in the frequency domain from a single optical parametric oscillator while in ref. \cite{menicucci11b} it was suggested to construct a time domain version of the QRL clusters state. With temporal encoding, the generated state has a cylindrical topology reminiscent of the DBSL, BSL and MBSL, allowing for computation along the cylinder with information encoded on the circumference of the cylinder. The scheme for generating the temporally encoded QRL state is summarized in Fig.~\ref{fig4:QRL}a. Since the QRL is self-inverse and bipartite, this QRL $\mathcal{H}$-graph state has a corresponding QRL cluster state (under phase rotations) which we consider in the following.

An efficient computation scheme on the QRL cluster state is presented in \cite{alexander16a} in the language of macronodes. It corresponds to the logic level marked on Fig.~\ref{fig4:QRL}a which is followed by a measurement device consisting of a beam-splitter network of four beam splitters ($\text{BS}_\text{1--4}$) and four homodyne detections. The cluster state at the logic level is shown in Fig.~\ref{fig4:QRL}b. With the logic level at a beam-splitter depth of only one, the edge weight of $t=\tanh(2r)$ is larger than in the DBSL, BSL and MBSL, while the self-loops are equal, $i\varepsilon=i\sech(2r)$. The logic cluster state consists of two-mode entangled states as in the generalized teleportation circuit in section \ref{sec2:gentele}, and no projection of the cluster state into wires before computation is necessary. This, together with the larger edge weight, reduces the gate noise and thus makes computation on the QRL more efficient. On the other hand, the increased complexity of the measurement device (a joint measurements of four modes) makes the computation scheme presented here more tricky and may seem less intuitive.

One computation step is marked in Fig.~\ref{fig4:QRL}b. It implements a two-mode operation from input modes $Ck$ (in1) and $Bk$ (in2) to the output modes $Ck+N$ (out1) and $Bk+1$ (out2). In the following, we will refer to the mode in computation from in1(2) to out1(2) as computation mode 1(2). 
It is possible to decouple the two computation modes 1 and 2 by restricting the basis settings to $\theta_{Ak}=\theta_{Dk}$ and $\theta_{Bk}=\theta_{Ck}$. In the same way as in Eq.~\eqref{eq3:compBS}, this effectively cancel the the beam splitter $\text{BS}_3$ and $\text{BS}_4$, since equal phase shifts commute with the beam splitter. Then, single-mode gates can be implemented using $\text{BS}_1$ and $\text{BS}_2$ in the same way as for the generalized teleportation in section \ref{sec2:gentele}, but the same gate will be applied to both computation modes 1 and 2 due to the basis restriction. That is, $\hat{U}\otimes\hat{U}$ will be implemented, where
\begin{equation}\label{eq4:U}
	\hat{U}=\Sq\left(\tanh(2r)\right)\R\left(\frac{\theta_+}{2}\right)\Sq\left(\tan\frac{\theta_-}{2}\right)\R\left(\frac{\theta_+}{2}\right)
\end{equation}
with $\theta_\pm=\theta_{Ck}\pm\theta_{Dk}$. Similarly, using the basis permutation rules in \cite{alexander16a}, restricting to $\theta_{Dk}=\theta_{Bk}$ and $\theta_{Ak}=\theta_{Ck}$ implements $\hat{U}\otimes\hat{U}$ on modes from in1 and in2 to out2 and out1, respectively. As a result, when implementing $\hat{U}\otimes\hat{U}$, the modes in computation may travel straight across each other on the cluster state lattice or they may do a $90^\circ$ change in their computation direction on the lattice depending on the basis restriction. For implementation of single mode gates, we will mainly focus on the former case in which one mode (computation mode 1) travels in the direction of the cluster state cylinder, while the other mode (computation mode 2), travelling around the cylinder, is ignored as illustrated in Fig.~\ref{fig4:QRL}c. Regardless of the basis restriction, the gate noise of one computation step is $\textbf{N}(\p_{Ak},\p_{Dk},\p_{Bk+1},\p_{Ck+N})^T$ with
\begin{equation*}
	\textbf{N}=\begin{pmatrix}
	-\frac{1}{\tanh(2r)}&0&0&0\\
	0&-\frac{1}{\tanh(2r)}&0&0\\
	0&0&1&0\\
	0&0&0&1
	\end{pmatrix}\;,
\end{equation*}
leading to equal quadrature noise factors in the two computation modes of
\begin{equation*}\begin{aligned}
	N_x&=\frac{1}{\tanh^2(2r)}\\
	N_p&=1
\end{aligned}\end{equation*}
in $\x$- and $\p$-quadratures, respectively.

As for the generalized teleportation circuit, the single-mode $\I$-gate is implemented in a single computation step with basis setting
\begin{equation*}
	\begin{pmatrix}
	\theta_+\\\theta_-
	\end{pmatrix}_I=
	\begin{pmatrix}
	0\\2\arctan\left(\tanh^{-1}(2r)\right)
	\end{pmatrix}
\end{equation*}
with gate noise variance $N_x\varepsilon/2$ and $N_p\varepsilon/2$ in $\x$- and $\p$-quadratures. The $\F$- and $\Ps(1)$-gates are implemented in two computation steps: With basis setting
\begin{equation*}
	\begin{pmatrix}
	\theta_{+1}\\\theta_{-1}\\\theta_{+2}\\\theta_{-2}
	\end{pmatrix}_F=\begin{pmatrix}
	\pi/2\\\pi/2\\0\\2\arctan\left(\tanh^{-2}(2r)\right)
	\end{pmatrix}\;
\end{equation*}
$\F$ is implemented with equal gate noise variance in $\x$ and $\p$ of $(N_x+N_p)\varepsilon/2$, while
\begin{equation*}
	\begin{pmatrix}
	\theta_{+1}\\\theta_{-1}\\\theta_{+2}\\\theta_{-2}
	\end{pmatrix}_P=\begin{pmatrix}
	\arctan2\\-\arctan2\\\pi/2\\\pi/2
	\end{pmatrix}
\end{equation*}
implements $\Ps(1)$ with gate noise variances of $2N_x\varepsilon/2$ and $2N_p\varepsilon/2$ in $\x$ and $\p$, respectively. Here, for the mode in computation travelling straight along the cylinder, $\theta_{\pm1}=\theta_{Ck}\pm\theta_{Dk}$ and $\theta_{\pm2}=\theta_{Ck+N}\pm\theta_{Dk+N}$ while $(\theta_{Bk},\theta_{Ak})=(\theta_{Ck},\theta_{Dk})$ and $(\theta_{Bk+N},\theta_{Ak+N})=(\theta_{Ck+N},\theta_{Dk+N})$.

To implement the $\CZ(1)$-gate, we have investigated $(\F^n\otimes\F^m)\CZ(1)$ for $n,m=0,1,2,3$ and find that $n=m=1$ leads to the lowest error probability in Eq.~\eqref{eq2:Perr} of the GKP-encoded qubits. With the basis setting
\begin{equation*}
	\begin{pmatrix}
		\theta_{Ak}\\\theta_{Bk}\\\theta_{Ck}\\\theta_{Dk}
	\end{pmatrix}_{C_Z}=
	\begin{pmatrix}
		\pi/2-\arctan(1/2)\\0\\\pi/2+\arctan(1/2)\\0
	\end{pmatrix}\;,
\end{equation*}
$(\Sq(t)\otimes\Sq(t))(\F\otimes\F)\CZ(1)$ is implemented in a single computation step, where the two modes in computation goes from in1 and in2 to out2 and out1 respectively (i.e. they do not cross, but each mode is redirected $90^\circ$). Here, $(\Sq(t)\otimes\Sq(t))$ is the distortion due to finite squeezing, and is compensated for in each computation mode in a second computation step with basis setting
\begin{equation*}
	\begin{pmatrix}
		\theta_+\\\theta_-
	\end{pmatrix}_{S(t^{-1})}=
	\begin{pmatrix}
		0\\2\arctan\left(\tanh^{-2}(2r)\right)
	\end{pmatrix}\;.
\end{equation*}
As a result, $(\F\otimes\F)\CZ(1)$ is implemented in two computation steps with equal gate noise variance in all four quadratures of $(N_x+N_p)\varepsilon/2$ as for the $\F$-gate. As gates on the QRL are in general performed on pairs of modes in computation and requires two computation steps (with the exception of the $\I$-gate), implementing $(\F\otimes\F)\CZ(1)$ among other computation modes may be tricky. However, an example of a possible implementation is shown in Fig.~\ref{fig4:QRL}d.

\begin{figure}
	\includegraphics[width=0.85\linewidth]{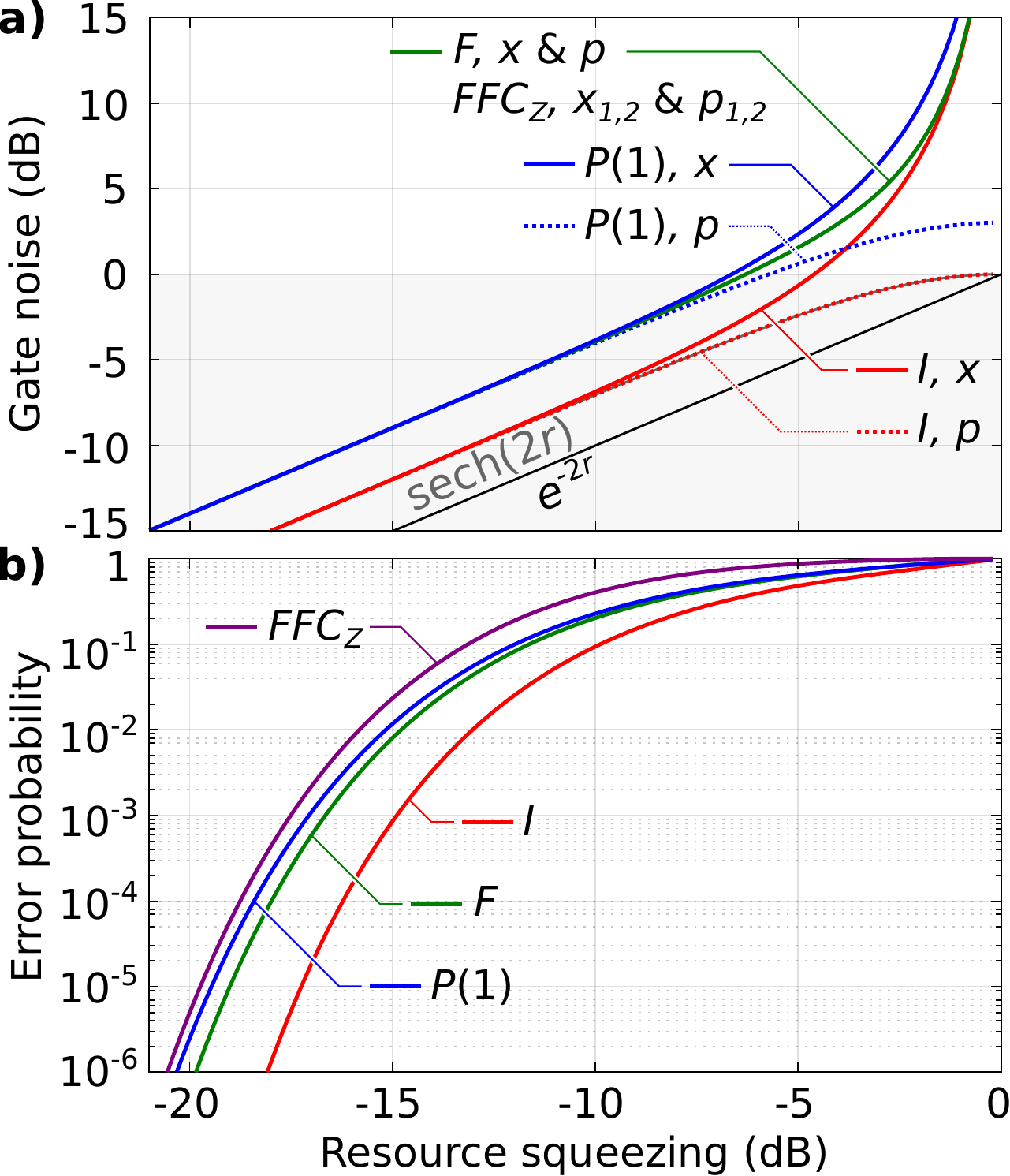}
	\caption{\label{fig4:QRL_result}(a) Gate noise variance of the $\I$-, $\F$-, $\Ps(1)$- and $(\F\otimes\F)\CZ(1)$-gate (in short $\F\F\CZ$) on the QRL cluster state as functions of input resource squeezing in Fig.~\ref{fig4:QRL}a. Here, $e^{-2r}$ and $\sech(2r)$ marks the resource and effective squeezing variance. Note that the gate noise variance in each of the four quadratures when implementing $\F\F\CZ$ is equal to the gate noise variance when implementing the $\F$-gate.  The gate noise here is shown together with the other considered schemes in appendix \ref{app:C} for easy comparison. (b) Resulting error probabilities of Eq.~\eqref{eq2:Perr} after quadrature corrections.}
\end{figure}

The gate noise variance for each of the implemented gates in $\lbrace\I,\F,\Ps(1),(\F\otimes\F)\CZ(1)\rbrace$ is shown in Fig.~\ref{fig4:QRL_result}a as a function of the initial squeezing of the $\p$-quadrature variance in the resource state, $e^{-2r}$. Notice that in the high squeezing limit, the gate noise of the $\I$-gate is equal to the effective variance squeezing of the cluster state modes, $\sech(2r)$, which is better than seen for the other computation schemes presented in this work, and is due to the large edge weight in the logic cluster state with no projection of the cluster state necessary before computation. The $\F$- and $\Ps(1)$-gate, implemented in two computation steps, naturally has double gate noise compared to the $\I$-gate, and so does the $(\F\otimes\F)\CZ(1)$-gate, unlike the $\CZ(1)$-gates implemented on the DBSL, BSL and MBSL. This improvement for the $(\F\otimes\F)\CZ(1)$-gate happens because no extra control modes are included when coupling two computation modes. In the limit of vanishing resource squeezing, the gate noise variance of each computation diverges in the $\x$-quadrature, erasing all information of the encoded state as for the generalized teleportation circuit in Eq.~\eqref{eq2:Wigner}.

To prevent gate noise accumulating on the GKP-encoded qubits, quadrature corrections as described in section \ref{sec2:errc} should be performed on modes in computation after each implemented gate. Here, with two computation modes in each computation step, two quadrature correction devices are necessary: One in each spatial mode $B$ and $C$ as marked in Fig.~\ref{fig4:QRL}a. After quadrature correcting modes as shown on the examples in Fig.~\ref{fig4:QRL}c and d, the error probabilities of Eq.~\eqref{eq2:Perr} are shown in Fig.~\ref{fig4:QRL_result}b for each of the four gates $\I$, $\F$, $\Ps(1)$ and $(\F\otimes\F)\CZ(1)$. As expected, with four successful quadrature corrections required to avoid qubit error, the $(\F\otimes\F)\CZ(1)$-gate has the highest error probability. In the following section \ref{sec5}, it is compared with the DBSL, BSL and MBSL.
	
\section{Discussion}\label{sec:5}
Below, in section \ref{sec5} we compare the cluster states and computation schemes presented and discussed in section \ref{sec:3} and \ref{sec:4}, while the figures summarizing the different computation schemes and resulting gate noise is shown side-by-side in appendix \ref{app:C}. In section \ref{sec:6} we then comment on computation universality with these cluster states.
	
\subsection{Cluster state comparison}\label{sec5}
\begin{figure}
	\includegraphics[width=0.85\linewidth]{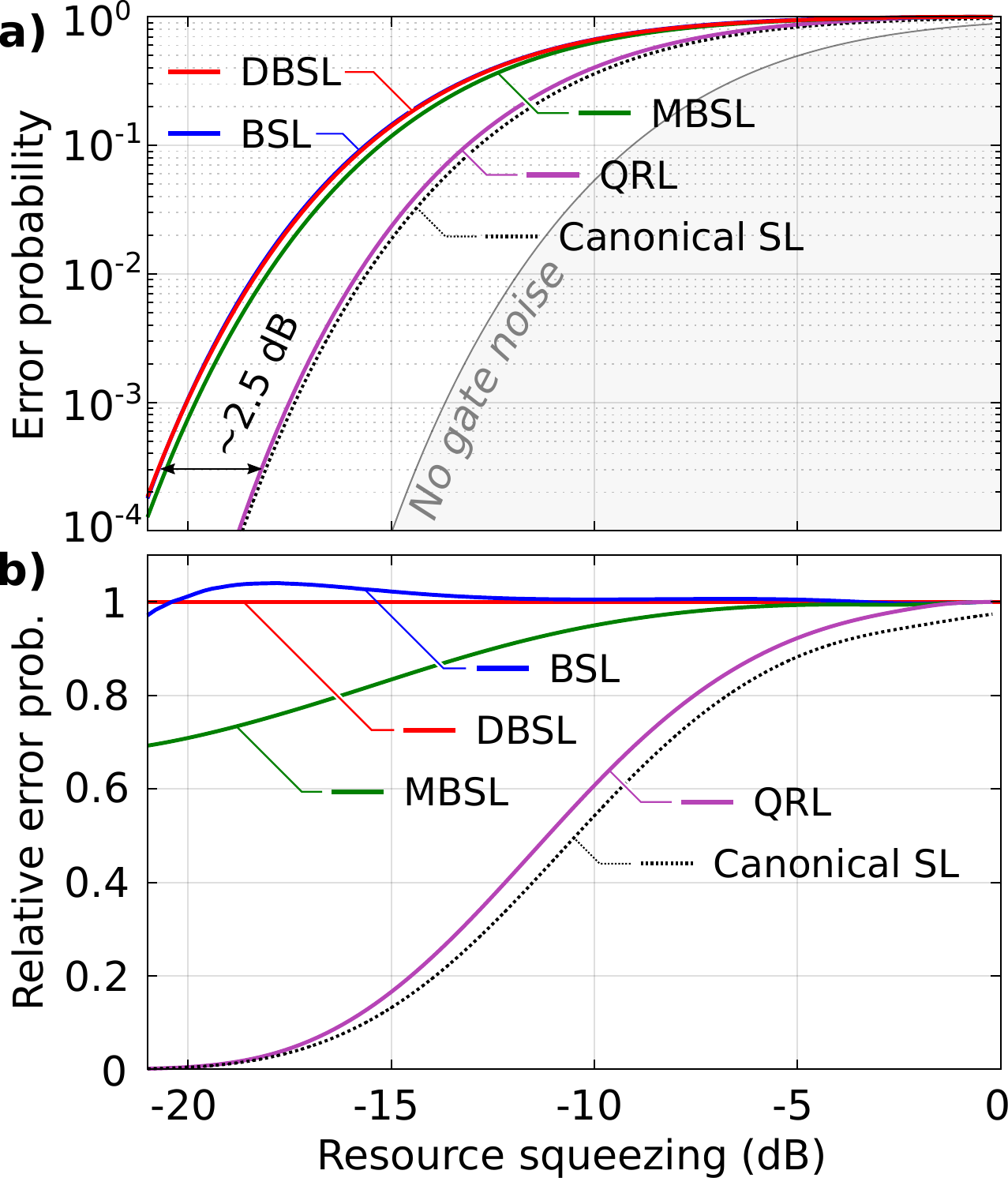}
	\caption{\label{fig5:comparison} (a) Error probabilities for $\CZ(1)$-gates implemented on the DBSL in section \ref{sec:3}, the BSL, MBSL and QRL in section \ref{sec:4}, and on the canonically generated square lattice (SL) cluster state in \cite{menicucci14}. Depending on the cluster state, the implemented $\CZ(1)$-gates have a Fourier gate bi-product on each mode. Note that the error probabilities for the DBSL and BSL are overlapping. The grey area marks the error probability in case of zero gate noise, where qubit errors are caused only by the available squeezing in the GKP-encoding. (b) Error probabilities in (a) relative to that of the DBSL.}
\end{figure}

For all the four cluster states considered in section \ref{sec:3} and \ref{sec:4}, the implemented two-mode $\CZ(1)$-gates lead to the highest error probability of the GKP-encoded qubits among the gates of the set $\lbrace\I,\F,\Ps(1),\CZ(1)\rbrace$. An indicative measure of the performance of a particular cluster state for quantum computing is thus the error probability associated with the implementation of the $\CZ(1)$-gate. In Fig.~\ref{fig5:comparison}a, these are plotted for the DBSL, BSL, MBSL and the QRL. Here, the error probability of $\CZ(1)$ implemented on a canonically generated square lattice (SL) cluster state in \cite{menicucci14} is plotted for comparison. 

As discussed in section \ref{sec2:errc}, the error probability in Eq.~\eqref{eq2:Perr} is fuelled by the gate noise, the noise of the GKP qubits as well as the noise introduced in quadrature error correction. Gate noise is governed by the amount of squeezing of cluster state while the noise of the qubits and correction is produced by the finite squeezing of the peaks in the GKP state. Here, as described in section \ref{sec3:noise_analysis}, we have assumed the peak variances of both quadratures in the GKP-states to equal the squeezing resource variance of $e^{-2r}/2$. To see how much the finite squeezing in the GKP-encoding and correction contributes to the error probability, the $\CZ(1)$ error probability in the case of zero gate noise (corresponding to setting $\boldsymbol{\sigma^2}=\textbf{0}$ in Eq.~\eqref{eq2:deltam}) is also plotted in Fig.~\ref{fig5:comparison}a. No matter what computation scheme is considered with the GKP-encoding used here, we will not be able to perform better than the case of zero gate noise, as the noise contributions from the GKP-encoding and quadrature correction are unavoidable.

The DBSL, BSL and MBSL are seen to have similar performance, while the QRL is superior and almost match the performance of the canonically generated SL cluster state. Approximately $\SI{2.5}{dB}$ additional squeezing is necessary in the squeezing resources for the DBSL and BSL to match the performance of the QRL. This performance advantage of the QRL is due to the larger cluster state edge weight in the logic level, and that the cluster state needs no projection by measurement of control modes, which adds additional noise to the state in computation. It is worth considering whether similar computation schemes can be developed for the DBSL, BSL and MBSL, possibly by placing the logic levels closer to the squeezing sources in the setups after the first beam-splitters (leading to temporally delocalized macronodes in the macronode language).

To quantify further the performance difference of the DBSL, BSL and MBSL, the $\CZ(1)$ error probabilities are plotted in Fig.~\ref{fig5:comparison}b relative to the $\CZ(1)$ error probability of the DBSL. Here, the BSL and DBSL are seen to have very similar performance in the investigated range of resource squeezing. The MBSL performs better with an error probability down to $70\%$ of the error probability in the DBSL at $\SI{21}{dB}$ resource squeezing, while the relative error probability is approximately $83\%$ using the currently achievable squeezing of  $\SI{15}{dB}$ \cite{vahlbruch16}. However, in practice one also has to account for experimental imperfections and setup complexity when deciding which setup to use: The generation scheme of the DBSL is technically simpler than that of MBSL as it requires only two squeezing sources and three interference points contra four squeezing sources and five interference points.

At first sight, with only two squeezing sources, the DBSL seems to require less resources than the BSL and MBSL. However, in the DBSL cluster state, only every second temporal mode holds control modes, while for the BSL and MBSL every single temporal mode includes both control \textit{and} wire modes. As a result, the DBSL only contains half as many modes for computation in the same number of temporal modes. Doubling the long delay, $N\tau$, in the generation setup in Fig.~\ref{fig3:DBSL}a doubles the cylindrical cluster state circumference and compensates for only having wire modes in every second temporal modes. However, by doubling the circumference, the time needed to implement gates doubles as well. As a result, there is a cost of using only two squeezing sources in the DBSL, which unfolds as fewer computation modes or longer computation time, but not as additional computation noise.

\begin{figure}
	\includegraphics[width=0.9\linewidth]{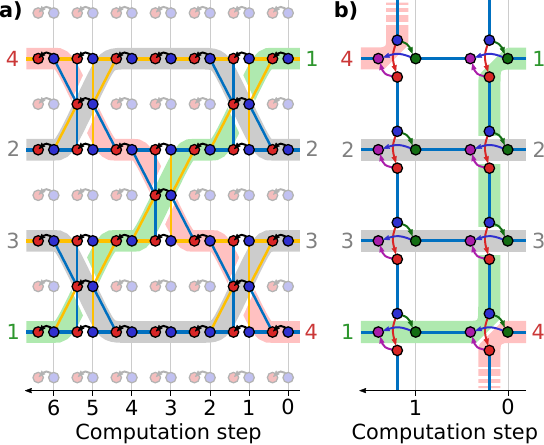}
	\caption{\label{fig5:swap}Implementation of a swap-gate between computation mode 1 and 4, $\hat{X}_{14}$, separated by computation mode 2 and 3, on (a) the DBSL and on (b) the QRL. Since only coupling between neighbouring wires is possible on the DBSL in (a), a circuit depth of 3 is required, corresponding to 6 horizontal computation steps  necessary along the cylindrical cluster state. The same goes for the BSL and the MBSL. On the QRL in (b), the same swap-gate can be implemented in a single horizontal computation step using crossing identity gates, $\I$. Here, computation mode 1 crosses computation mode 2 and 3, while computation mode 4 is lead all the way around the cluster state cylinder to appear in the next horizontal computation step, crossing other computation modes on its way.}
\end{figure}

Finally, we compare the architecture of the computation schemes on the considered cluster states. The DBSL, BSL and MBSL all use the same principles of measuring control modes to control coupling between wires with modes in computation. Turning on and off coupling between wires makes it intuitive to implement multi-mode gates decomposed into single- and two-mode gates, while on the QRL one has to take care of the surrounding modes when implementing two-mode gates as for the $\CZ(1)$-gate in Fig.~\ref{fig4:QRL}d. However, the control-mode based architectures only allow coupling between neighbouring wires, whereas the QRL is more ``flexible'' as introduced in \cite{alexander16a}. As an example, consider an arbitrary swap-gate, $\hat{X}_{ij}$, swapping the modes in computation on wires $i$ and $j$. On the DBSL, a swap-gate (with an unimportant Fourier gate applied to the two output modes) can be performed between two neighbouring wires in two computation steps from temporal mode $(k-2,k)$ to $(k+2N,k+2N-2)$ with the basis setting
\begin{equation*}
	\begin{pmatrix}
		\theta_{Ak-2}\\
		\theta_{Bk-2}\\
		\theta_{Ak}\\
		\theta_{Bk}\\
		\theta_{Ak+N-2}\\
		\theta_{Bk+N-2}\\
		\theta_{Ak+N-1}\\
		\theta_{Bk+N-1}\\
		\theta_{Ak+N}\\
		\theta_{Bk+N}
	\end{pmatrix}_X=
	\begin{pmatrix}
		\pi/4\\
		-\pi/4\\
		\pi/4\\
		-\pi/4\\
		\pi/2\\
		0\\
		\pi/2\\
		0\\
		\pi/2\\
		0
	\end{pmatrix}
\end{equation*}
independent on the amount of resource squeezing, and with a gate noise variance of $N_x\varepsilon/2$ and $N_p\varepsilon/2$ in $\x$- and $\p$-quadratures, respectively, where $N_x=\tanh^{-4}(2r)+3\tanh^{-2}(2r)$ and $N_p=\tanh^2(2r)+3$. Thus, to swap two modes on wires separated by $n$ wires in-between, $n+1$ swap-gates are required on each mode leading to $2(n+1)$ required computation steps. On the QRL, on the other hand, using vertically travelling modes in Fig.~\ref{fig4:QRL}, two modes can be swapped in only a single horizontal computation step independent on the initial distance between the modes on the cluster state lattice. This is illustrated in Fig.~\ref{fig5:swap} for $\hat{X}_{14}$ with a mode distance of 2. As a result, depending on the interconnectivity required in the quantum algorithm to be implemented, computation times can be shorter on the QRL than on the DBSL, BSL and MBSL.

Lastly, we want to comment on the performance of the QRL compared to the canonically generated SL cluster state. It is clear that the QRL performs almost as well as the SL, while the SL is much more challenging to generate since it requires on-line squeezing to perform canonical $\CZ(g)$ operations and the total squeezing cost is in general larger \cite{gu09,braunstein05}. However, for a fair comparison, it should be mentioned that the $\CZ(1)$ implemented in \cite{menicucci14} on the SL was not optimised. It was implemented with four computation steps, where each of the $\x$- and $\p$-quadrature corrections in the GKP-scheme were performed in two different computation steps, both leading to more noise on the GKP-encoded qubits. The $\CZ(1)$ error probability on the SL may be improved by optimizing the required basis settings to implement the $\CZ(1)$-gate in fewer computation steps, and performing GKP quadrature corrections of both the $\x$- and $\p$-quadrature on the last cluster state mode as for the computation schemes considered in this work.

\subsection{Towards universality and fault-tolerance}\label{sec:6}
The four different computation schemes in section \ref{sec:3} and \ref{sec:4} involve only Gaussian measurements (in the form of homodyne detection) on Gaussian cluster states. In this pure Gaussian realm, one is only able to perform universal Gaussian computation \cite{gu09}, which with Gaussian input-modes may be simulated classically \cite{bartlett02a,bartlett02b}. To achieve universal quantum computation, non-Gaussian operations or resources are required \cite{lloyd99}. There exist different proposals on how to achieve a universal gate set, which we summarize and discuss in the following.
Non-Gaussianity of states and operations has been obtained in numerous systems \cite{andersen15}, including some recent results on optical non-Gaussian state preparation \cite{hacker19} and non-Gaussian transformations on cluster states \cite{ra20}.

In many CV quantum computing architecture proposals, the Gaussian gate set is complemented with the non-Gaussian cubic phase gate, $\hat{K}(\chi)=e^{i\chi\x^3/3}$ \cite{gottesman01,gu09} to achieve universal quantum computation on the bosonic modes \cite{lloyd99}. Such a non-Gaussian gate can for example be implemented by redirecting specific modes of the cluster to a photon counter, thereby realizing a measurement induced non-Gaussian gate transformation \cite{gu09}. Moreover, in \cite{alexander18} it is shown how $\hat{K}(\chi)$ may be implemented on the BSL using an ancillary cubic phase state, $\ket{\chi}=\int \text{d}s\,e^{i\chi s^3/3}\ket{s}_x$, as a non-Gaussian resource switched into the logic level of the computation scheme as an input state. Such cubic phase state may be prepared using photon counting \cite{gottesman01}. Given the similarities between computation on the BSL, the DBSL and the MBSL, it is straight forward to adopt this method of implementing $\hat{K}(\chi)$ by inputting $\ket{\chi}$ in these computation schemes. A similar approach may also be viable on the QRL.

Using GKP-states with symmetric quadrature noise, one can expect bad performance of the cubic phase gate due to the applied phase by $\hat{K}(\chi)$ on the finitely squeezed GKP peaks being a cubic function of $\x$ \cite{gottesman01,hastrup20}. A more efficient approach to quantum universality is to consider a gate set that is only universal in the encoded logic space rather than in the full, infinite-dimensional Hilbert space. This requires an ample supply of qubit magic states such as the Hadamard eigenstates, $\ket{H_L}=\cos\pi/8\ket{0_L}+\sin\pi/8\ket{1_L}$. By injecting these states into the computation wires as input states using an optical switch, the non-Clifford $\pi/8$-gate can be executed with only Gaussian transformations of the bosonic modes \cite{gottesman01,yamasaki19}. Such magic GKP states may be prepared similarly to the GKP-encoded input states, or directly distilled using GKP $\ket{0_L}$-states \cite{baragiola19}. In conclusion, the inherent non-Gaussianity of the GKP-states is sufficient to achieve universal quantum computation in the GKP-encoded qubit subspace using solely Gaussian transformations. Moreover, adding magic state distillation to the scheme may not increase the experimental requirements significantly since the squeezing needed for the distillation is expected to be lower than the squeezing already required to reach fault-tolerant Clifford computation \cite{menicucci14,baragiola19}.

Finally, for fault-tolerant computation, the qubit error correction scheme---concatenated with the GKP error correction scheme---should be considered when estimating the required squeezing. With the quadrature corrections of states in computation after each implemented gate, gate noise and finite squeezing in the approximate GKP-encoded qubit states are translated to qubit errors. For fault-tolerant computation, these qubit errors are corrected with an appropriate qubit error correction scheme, where a logic qubit is encoded in multiple GKP-qubits. Here, it is not appropriate just to choose a qubit error correction scheme with a large qubit error threshold, as considerations on how to practically implement the scheme are also of critical importance. For this reason, here we will not estimate a squeezing threshold for fault-tolerance. As an example, the 7-qubit Steane code with a $\sim10^{-3}$ error threshold requires two-mode gates between arbitrary modes in computation \cite{steane03}, while the discussed computation schemes in section \ref{sec:3} and \ref{sec:4} only implement two-mode gates between neighbouring computation modes. Thus, to implement the 7-qubit Steane code a number of swap-gates are required for each syndrome measurement, each leading to an increase in the combined qubit error probability before qubit error correction. The QRL may have an advantage when considering the implementation of a qubit error correction scheme owing to its flexibility as previously discussed and illustrated for a swap-gate in Fig.~\ref{fig5:swap}. Future work includes considerations on the practical implementation of qubit error correction on a suitable and realizable cluster states.

\section{Conclusion}\label{sec:conclusion}
In summary, we have reviewed the principles of CV measurement-based QC based on generalized teleportation, we have proposed an efficient computation scheme for the DBSL cluster state that was experimentally generated in \cite{larsen19}, and we have carefully analysed and compared quantum computation based on that state with the BSL, QRL, and MBSL cluster states.

Through a careful study of the added gate noise for the different cluster states, we find that the DBSL, the BSL and the MBSL exhibit similar performance. We also find that the QRL is superior in terms of performance and flexibility, allowing implementation of quantum circuits in a minimum number of time steps. Finally, we have reviewed proposals for implementation of a universal gate set, either on the bosonic modes or just in the GKP-encoded qubit subspace, and conclude that universal qubit computation is possible in all four considered cluster states, given the availability of GKP-states.

To optimise the performance of the various computation schemes, we introduced a tool to find the basis setting implementing a desired gate with minimum GKP-encoded qubit errors. We believe that this technique for finding the optimal basis settings will be important for future developments and optimizations of new types of gates and algorithms. It should however be noted that the technique of optimizing the basis setting might not be the only strategy for minimizing the error probability: We have only considered GKP-qubit encodings on a square grid in phase space which is appropriate for symmetric noise addition among conjugate quadratures. However, since the considered computation schemes in general add noise asymmetrically in the quadratures, it may be beneficial to encode the qubits in a rectangular lattice. Since different gates have different gate noise asymmetry in $\x$- and $\p$-quadrature, the optimal lattice ratio depends on which gates dominate the circuit to be implemented: As an example, the Fourier gate, $\F$, in general adds symmetric gate noise in which case a square lattice is optimal, while the gate noise asymmetry of the identity gate, $\I$, depends on the resource squeezing. One can argue, that with the $\hat{C}_Z$-gate being the noisiest gate, the GKP lattice ratio should be optimized to minimize qubit error for this gate. In this case, for the DBSL, BSL and MBSL, the optimal ratio again depends on the resource squeezing, while for the superior QRL with symmetric gate noise, the square lattice seems optimal. However, one further complication is that when performing the $\hat{C}_Z$-gate or the $\hat{P}$-gate, not only gate noise is added, but also noise from the state in computation is added due to the addition of quadratures in these gates. Thus, the optimal lattice ratio depends as well on the noise performance of the states in computation, and to determine a general optimal lattice ratio for a given application is outside the scope of this work. Finally, one has to keep in mind that changing the lattice ratio also alters the logic operators in the GKP-encoding. As an example, with a rectangular lattice the logic Hadamard gate becomes a combination of the Fourier and squeezing gates.

Throughout this article, we have assumed all cluster states to be pure, while in practice, the cluster state will have some degree of mixedness in the form of excess noise in the anti-squeezed quadratures. However, it has been shown in ref. \cite{walshe19} that excess noise in the anti-squeezed quadrature does not affect the performance of the computation, and thus our purity assumption in this article is well justified. It is however worth mentioning that in practice it is still favorable to produce highly pure squeezed states as large excess noise will decrease the amount of squeezing due to inevitable phase instabilities of the experimental setup. 

In this article, we have not studied the actual implementation of qubit error correction. Thus, as an outlook, it would be interesting to study how a qubit error correction algorithm is most efficiently implemented such that the squeezing threshold for fault-tolerant quantum computation is minimized. An interesting solution could be topological QC, for which the resulting squeezing threshold is within the already experimental demonstrated range \cite{fukui18,noh20}. In topological QC, the qubits are encoded in a two-dimensional plane while the actual computation takes place in a third dimension, thus rendering the need for the construction of 3D cluster states. Proposals do exist for the generation of 3D cluster states \cite{wu19,fukui20}, and the next interesting step is thus to analyse the performance of these states using the techniques developed in this article. 

\begin{acknowledgments}
	We acknowledge useful discussion with Rafael N. Alexander and Peter van Loock. The work was supported by the Danish National Research Foundation through the Center for Macroscopic Quantum States (bigQ, DNRF0142).
\end{acknowledgments}

\appendix

\section{Calculation of quadrature tansformations}\label{app:A}
\begin{figure*}
	\includegraphics[width=\linewidth]{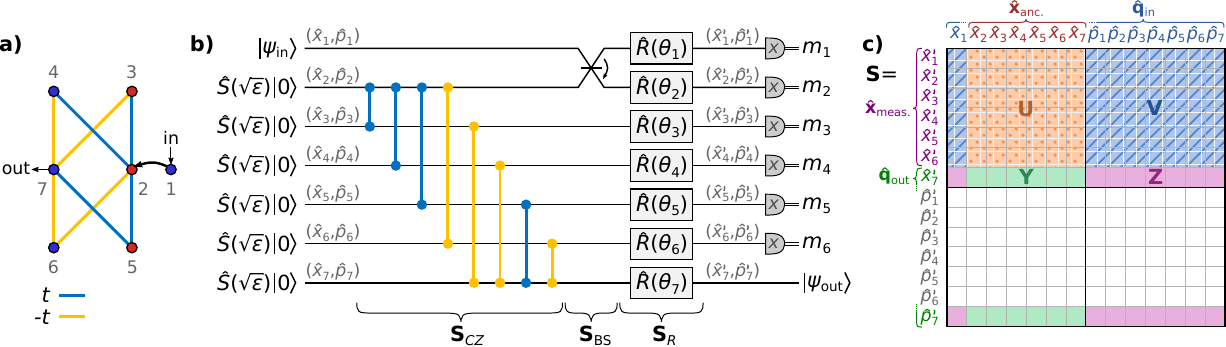}
	\caption{\label{figA:circuit} (a) Graph notation of a single-mode computation step on the DBSL. For an input in temporal mode $k$, the mode numbering translates as $(1,2,3,4,5,6,7)=(Bk,Ak,Ak-1,Bk+N-1,Ak+1,Bk+N+1,Bk+N)$ where $A$ and $B$ are spatial modes in Fig.~\eqref{fig3:DBSL}a. (b) Corresponding circuit where the blue and yellow two-mode gates represent $\CZ(g)$-gates of weights $t$ and $-t$, respectively. (c) Outline of the symplectic matrix $\textbf{S}=\textbf{S}_{R}\textbf{S}_{BS}\textbf{S}_{CZ}$ representing the quadrature transformation in (b).}
\end{figure*}

In this appendix, we present an example of the quadrature transformation of the single-mode computation step on the DBSL that leads to the expressions Eq.~\eqref{eq3:Singlemode} and \eqref{eq3:INF} in section \ref{sec:3}. The modes involved are shown on the graph in Fig.~\ref{figA:circuit}a with the corresponding circuit in Fig.~\ref{figA:circuit}b. We will use the mode numbering labelled in Fig.~\ref{figA:circuit}a.

The approximate cluster state (ancillary mode 2--7) consists of vacuum states squeezed by $\sqrt{\varepsilon}$ and connected by $\CZ(g)$-operations of weights that are described by the adjacency matrix
\begin{equation*}
	\textbf{A}=\begin{pmatrix}
		0 & 0 & 0 & 0 & 0 & 0 & 0\\
		0 & 0 & t & t & t & -t & 0\\
		0 & t & 0 & 0 & 0 & 0 & -t\\
		0 & t & 0 & 0 & 0 & 0 & -t\\
		0 & t & 0 & 0 & 0 & 0 & t\\
		0 & t & 0 & 0 & 0 & 0 & -t\\
		0 & 0 & -t & -t & t & -t & 0\\
	\end{pmatrix}\;.
\end{equation*}
Thus, in the Heisenberg picture, we consider the generation of the cluster state as a quadrature transformation described by the symplectic matrix
\begin{equation*}
	\textbf{S}_{CZ}=\begin{pmatrix}
		\textbf{I} & \textbf{0}\\ \textbf{A} & \textbf{I}
	\end{pmatrix}
\end{equation*}
on the input and initially squeezed ancillary modes, where \textbf{I} and \textbf{0} are the $7\times7$ identity and zero matrix, respectively (note that the quadratures of the input mode 1 are left unchanged by $\textbf{S}_{CZ}$). The input mode 1 is then connected to the cluster state by a beam-splitter (the beam-splitter of the measurement device in Fig.~\ref{fig3:DBSL}a), leading to the quadrature transformation
\begin{equation*}
	\textbf{S}_{BS}=\begin{pmatrix}
		\textbf{B} & \textbf{0}\\
		\textbf{0} & \textbf{B}
	\end{pmatrix}\;,
\end{equation*}
where
\begin{equation*}
	\textbf{B}=\begin{pmatrix}
	1/\sqrt{2} & -1/\sqrt{2} & 0 & 0 & 0 & 0 & 0\\
	1/\sqrt{2} & 1/\sqrt{2} & 0 & 0 & 0 & 0 & 0\\
	0 & 0 & 1 & 0 & 0 & 0 & 0\\
	0 & 0 & 0 & 1 & 0 & 0 & 0\\
	0 & 0 & 0 & 0 & 1 & 0 & 0\\
	0 & 0 & 0 & 0 & 0 & 1 & 0\\
	0 & 0 & 0 & 0 & 0 & 0 & 1
	\end{pmatrix}\;.
\end{equation*}
Here we have ignored beam-splitter operations on the control modes 3--6, as these operations can be directly compensated for by adding/subtracting the measurement outcomes of the homodyne detectors as shown in Eq.~\eqref{eq3:compBS}. Finally, each mode $j$, except the output mode 7, is measured in basis $\x(\theta_j)$. This is represented first by a phase rotation, $\R(\theta_j)$, followed by a homodyne measurement of $\x_j$. Thus, before the $\x$-measurements, the quadratures are transformed as
\begin{equation*}
	\textbf{S}_R=\begin{pmatrix}
	\textbf{c} & \textbf{s}\\-\textbf{s}&\textbf{c}
	\end{pmatrix}\;,
\end{equation*}
where $\textbf{c}$ and $\textbf{s}$ are matrices with $(\cos\theta_1,\cdots,\cos\theta_7)$ and $(\sin\theta_1,\cdots,\sin\theta_7)$ in the diagonal, respectively, and zero elsewhere. For implementing single-mode gates, the control mode measurement bases are set to $(\theta_3,\theta_4,\theta_5,\theta_6)=(-1)^i\theta_c(1,1,-1,-1)$, where $\theta_c=\pi/4$ for simplicity, and the output mode cannot be phase rotated, $\theta_7=0$. Here, $i$ is the wire number as shown in Fig.~\ref{fig3:DBSL}b. The total quadrature transformation of the input and squeezed cluster state modes just before measurements is then
\begin{equation*}
	\hat{\textbf{q}}'=\textbf{S}_R\textbf{S}_{BS}\textbf{S}_{CZ}\hat{\textbf{q}}=\textbf{S}\hat{\textbf{q}}\;,
\end{equation*}
where $\hat{\textbf{q}}=(\x_1,\cdots,\x_7,\p_1,\cdots,\p_7)^T$ and $\hat{\textbf{q}}'=(\x'_1,\cdots,\x'_7,\p'_1,\cdots,\p'_7)^T$ are vectors of quadrature operators before and after the transformation as marked in Fig.~\ref{figA:circuit}b. It should be noted that for the cluster state prepared as a $\mathcal{H}$-graph with off-line squeezing only, the effective amount of squeezing of the cluster state modes is $\varepsilon=\sech(2r)$, where $r$ is the squeezing parameter of the initially prepared off-line squeezed state with variance $e^{-2r}$ \cite{gu09}. 

Next, we solve for the anti-squeezed $\x$-quadratures of the cluster state modes, $\hat{\textbf{x}}_\text{anc.}=(\x_2,\cdots,\x_7)^T$, as a function of the measured $\x$-quadratures, $\hat{\textbf{x}}_\text{meas.}=(\x'_1,\cdots,\x'_6)^T$:
\begin{equation*}
	\hat{\textbf{x}}_\text{meas.}=\textbf{U}\hat{\textbf{x}}_\text{anc.}+\textbf{V}\hat{\textbf{q}}_\text{in}
\end{equation*}
\begin{equation*}
\Updownarrow
\end{equation*}
\begin{equation*}
	\hat{\textbf{x}}_\text{anc.}=\textbf{U}^{-1}\hat{\textbf{x}}_\text{meas.}-\textbf{U}^{-1}\textbf{V}\hat{\textbf{q}}_\text{in}\;,
\end{equation*}
where $\hat{\textbf{q}}_\text{in}=(\x_1,\p_1,\cdots,\p_7)^T$ while $\textbf{U}$ and $\textbf{V}$ are the parts of $\textbf{S}$ that transform $\hat{\textbf{x}}_\text{anc.}$ and $\hat{\textbf{q}}_\text{in}$ to $\hat{\textbf{x}}_\text{meas.}$ as shown in Fig.~\ref{figA:circuit}c. Finally, we substitute the $\x$-quadratures of the cluster state with the quadratures of output mode 7, $\hat{\textbf{q}}_\text{out}=(\x'_7,\p'_7)$:
\begin{equation*}\begin{aligned}
	\hat{\textbf{q}}_\text{out}&=\textbf{Y}\hat{\textbf{x}}_\text{anc.}+\textbf{Z}\hat{\textbf{q}}_\text{in}\\
	&=\textbf{Y}\left(\textbf{U}^{-1}\hat{\textbf{x}}_\text{meas.}-\textbf{U}^{-1}\textbf{V}\hat{\textbf{q}}_\text{in}\right)+\textbf{Z}\hat{\textbf{q}}_\text{in}\\
	&=\left(\textbf{Z}-\textbf{Y}\textbf{U}^{-1}\textbf{V}\right)\hat{\textbf{q}}_\text{in}+\textbf{Y}\textbf{U}^{-1}\hat{\textbf{x}}_\text{meas.}\;.
\end{aligned}\end{equation*}
With $\hat{\textbf{x}}_\text{meas.}\rightarrow(m_1,\cdots,m_6)$ when measuring, $\textbf{Y}\textbf{U}^{-1}\hat{\textbf{x}}_\text{meas.}$ corresponds to the by-product displacement, while $\textbf{M}\equiv\textbf{Z}-\textbf{Y}\textbf{U}^{-1}\textbf{V}$ of size $2\times8$ corresponds to the combined gate symplectic matrix $\textbf{G}$ and gate noise matrix $\textbf{N}$ in Eq.~\eqref{eq2:trans} as $\textbf{M}=(\textbf{G} \; \textbf{N})$. Extracting $\textbf{G}$ as the first two columns of $\textbf{M}$ transforming $(\x_1,\p_1)$ to $(\x'_7,\p'_7)$, we get
\begin{equation*}
	\textbf{G}=\frac{1}{\sin\theta_-}\begin{pmatrix}
		\frac{1}{t'}\cos\theta_++\frac{1}{t'}\cos\theta_- & \frac{1}{t'}\sin\theta_+\\
		-t'\sin\theta_+ & t'\cos\theta_+-t'\cos\theta_-
	\end{pmatrix}
\end{equation*}
which is the symplectic matrix corresponding to the operation in Eq.~\eqref{eq3:Singlemode} where $t'=(-1)^i4t^2$ and $\theta_\pm=\theta_1\pm\theta_2$. $\textbf{N}$ is associated with the remaining 6 columns of $\textbf{M}$;
\begin{equation}\label{eqA:N}
	\textbf{N}=\begin{pmatrix}
		-\frac{1}{4t^2} & \frac{1}{4t} & \frac{1}{4t} & -\frac{1}{4t} & \frac{1}{4t} & 0\\
		0 & t & t & t & -t & 1
	\end{pmatrix}\;,
\end{equation}
which leads to the quadrature noise factors coined in Eq.~\eqref{eq3:INF} when $t=\tanh(2r)/2$.

The procedure shown here for calculating the gate symplectic matrix, $\textbf{G}$, and gate noise matrix, $\textbf{N}$, is not limited to the single-mode one computation step on the DBSL, but represents a general procedure that can be used to analyse the noise of all gates in this work (irrespective of the cluster state) as done in section \ref{sec:3} and \ref{sec:4}: If $\textbf{S}_{CZ}$ represents the construction of any cluster state and $\textbf{S}_R\textbf{S}_{BS}$ represents any Gaussian measurement, we can determine the resulting linear quadrature transformation corresponding to an arbitrary Gaussian operation on a single- or multi-mode input state. For each case, we need to keep track of the following quadratures: $\hat{\textbf{x}}_\text{anc.}$ including anti-squeezed $\x$-quadratures of the cluster state; $\hat{\textbf{q}}_\text{in}$ including the input mode quadratures and squeezed $\p$-quadratures of the cluster state leading to gate noise; $\hat{\textbf{x}}_\text{meas.}$ including the transformed $\x$-quadratures to be measured; and $\hat{\textbf{q}}_\text{out}$ including the output mode quadratures of non-measured modes.

\begin{widetext}
\section{Wigner function transformations}\label{app:B}
In this appendix we discuss the single-mode computation step in the DBSL, BSL and MBSL in the Wigner function representation. Here, for simplicity, the basis setting for implementing the $\I$-gate is chosen, while to shorten the notation, we have post-selected on measurement outcomes equal zero. As described in the main text section \ref{sec2:gentele}, non-zero measurement outcomes lead to an unimportant displacement in phase-space. For the QRL, the two-mode cluster state corresponds to the one considered for the generalized teleportation in section \ref{sec2:gentele}, and so the Wigner function transformation is similar to that presented in Eq.~\eqref{eq2:Wigner}.

For a single-mode $\I$-gate performed on the DBSL in one computation step, described in sections \ref{sec:3} A and B, the transformation of the Wigner function can be calculated in the same way as we did for the generalized teleportation in section \ref{sec2:gentele}, resulting in
\begin{equation*}\begin{aligned}
	W_\text{out}(x,p)=&\;\mathcal{N}G_{1/\varepsilon}(x)\int\text{d}\eta_4\,G_\varepsilon(\eta_4)G_{4t^2/\varepsilon}(p-\eta_4)\int\text{d}\eta_3\,G_{\varepsilon/(4t^2)}(\eta_3)\\
	&\quad\quad G_{1/(4t^2\varepsilon)}(x-\eta_3)\int\text{d}\eta_2\,G_{4t^2\varepsilon}(\eta_2)G_{16t^4/\varepsilon}(p-\eta_2-\eta_4)\int\text{d}\eta_1\,G_{\varepsilon/(16t^4)}(\eta_1)W_\text{in}\begin{pmatrix}
	x-\eta_1-\eta_3\\p-\eta_2-\eta_4
	\end{pmatrix}\;,
\end{aligned}\end{equation*}
where $\mathcal{N}$ is a normalization factor and $G_\delta$ is a normalized Gaussian function of $\delta/2$ variance. The transformation includes two convolutions in each quadrature and corresponding envelopes in the conjugate quadrature due to the Fourier relation between quadratures. Comparing with Eq.~\eqref{eqA:N}, and referring to the mode numbering in Fig.~\ref{figA:circuit}a, in the $\x$-quadrature the first convolution with $G_{\varepsilon/(16t^4)}$ corresponds to noise from the finitely squeezed mode 2, while the third convolution with $G_{\varepsilon/(4t^2)}$ corresponds to noise of control modes 3, 4, 5 and 6. In the $\p$-quadrature, the second convolution with $G_{4t^2\varepsilon}$ corresponds to noise from the control modes 3, 4, 5 and 6, while the last convolution with $G_{\varepsilon}$ corresponds to the finite squeezing noise of the output mode 7. In the limit of infinite squeezing, $r\rightarrow\infty$, (assuming $t\neq0$) the convolution functions become delta functions since $\varepsilon=\sech(2r)\rightarrow0$, while their corresponding envelopes in the orthogonal quadratures become infinitely broad, and so $W_\text{out}(x,p)\rightarrow W_\text{in}(x,p)$. In the limit of $t=0$ where we expect no information to pass from the input mode 1 to the output mode 7, the first three convolutions lead to the Wigner function of an infinitely squeezed state in $\p$, erasing all information of the input state, while the last convolution with $G_\varepsilon$ in $\p$-quadrature ensures that the output Wigner function equals the initial squeezed Wigner function of mode 7, $W_\text{out}(x,p)=G_{1/\varepsilon}(x)G_\varepsilon(p)$, which equals vacuum for no squeezing as $\varepsilon=\sech(2r)=1$ when $r=0$.

On the BSL, the single-mode $\I$-gate performed in one computation step transforms the Wigner function as
\begin{equation*}\begin{aligned}
	W_\text{out}(x,p)=&\;\mathcal{N}
	G_{1/\varepsilon}(x)\int\text{d}\eta_4\,G_\varepsilon(\eta_4)
	G_{2t^2/\varepsilon}(p-\eta_4)\int\text{d}\eta_3\,G_{\varepsilon/(2t^2)}(\eta_3)\\
	&\quad\quad G_{1/(2t^2\varepsilon)}(x-\eta_3)\int\text{d}\eta_2\,G_{2t^2\varepsilon}(\eta_2)G_{4t^2/\varepsilon}(p-\eta_2-\eta_4)\int\text{d}\eta_1\,G_{\varepsilon/(4t^4)}(\eta_1)W_\text{in}\begin{pmatrix}
	x-\eta_1-\eta_3\\p-\eta_2-\eta_4
	\end{pmatrix}\;.
\end{aligned}\end{equation*}
Similar to the DBSL, comparing with $\textbf{N}$ in Eq.~\eqref{eq4:BSL_N}, the convolutions with $G_{\varepsilon/(4t^4)}$ and $G_{\varepsilon/(2t^2)}$ in the $\x$-quadrature correspond respectively to noise added from the first wire mode $Ak$ and the two control modes $Bk+1$ and $Ck+N$ in a square cluster of the BSL in Fig.~\ref{fig4:BSL}a,b. In the $\p$-quadrature, the convolutions with $G_{2t^2\varepsilon}$ corresponds to noise from the control modes $Bk+1$ and $Ck+N$, while the convolution with $G_{\varepsilon}$ corresponds to noise from the output mode $Dk+N$. Again, in the limit of infinite squeezing, $W_\text{out}(x,p)\rightarrow W_\text{in}(x,p)$, while for $t=0$ the output Wigner function becomes $W_\text{out}(x,p)=G_{1/\varepsilon}(x)G_\varepsilon(p)$ as expected.

For the MBSL, the Wigner function transformation of the $\I$-gate in one computation step with the control basis $\theta_c=\pi/2$ is
\begin{equation*}\begin{aligned}
	W_\text{out}(x,p)=&\;\mathcal{N}G_{1/\varepsilon}(x)G_{2\varepsilon+4t^2/\varepsilon}(p)\int\text{d}\eta_3\,G_{\varepsilon/(4t^2)}(\eta_3)
	G_{1/\varepsilon}(x-2\eta_3)\int\text{d}\eta_2\,G_{2\varepsilon}(\eta_2)\\
	&\quad\quad G_{4t^2/\varepsilon}(p-\eta_2)\int\text{d}\eta_1\,G_{\varepsilon/(4t^2)}(\eta_1)
	W_\text{in}\begin{pmatrix}
	x-\eta_1-\eta_3\\p-\eta_2
	\end{pmatrix}\;.
\end{aligned}\end{equation*}
Due to the direct edges along the computation wires of the butterfly cluster states in Fig.~\ref{fig4:Asavanant}, the Wigner function transformation becomes less intuitive. Here, the envelope $G_{2\varepsilon+4t^2/\varepsilon}(p)$ corresponds to an envelope of $G_{4t^2/\varepsilon}(p)$ convoluted with $G_{2\varepsilon}$. Comparing with Eq.~\eqref{eq4:A_N}, the first and third convolution in the $\x$-quadrature, both with $G_{\varepsilon/(4t^2)}$, correspond to noise added from wire mode $Dk$ and control mode $Ak$. The second convolution with $G_{2\varepsilon}$ in the $\p$-quadrature corresponds to noise from both control mode $Bk+1$ and the output mode $Ck+N$. In the infinite squeezing limit, $W_\text{out}(x,p)\rightarrow W_\text{in}(x,p)$. For $t=0$ we get $W_\text{out}(x,p)=G_{1/\varepsilon}(x)G_{2\varepsilon}(p)G_{2\varepsilon}(p)=G_{1/\varepsilon}(x)G_\varepsilon(p)$.

If, instead of the identity gate $\I$, an arbitrary single-mode Gaussian gate of one computation step is implemented with symplectic matrix $\textbf{G}$, the resulting Wigner function transformation corresponds to that presented above, but with the arguments of $W_\text{in}$ transformed by $\textbf{G}^{-1}$ as shown in Eq.~\eqref{eq2:Wigner} for the generalized teleportation. For single-mode gates implemented in two computation steps, the output Wigner function of the first step becomes the input Wigner function of the second step, leading to addition of the gate noise variance since a convolution of two Gaussian functions is a Gaussian function with the combined variance, i.e. additive Gaussian gate noise. For multi-mode gates, more modes are involved leading to more convolutions in the expression of the output Wigner function, and the Wigner function representation becomes tedious. However, the principle is the same as for single-mode gates: The gate noise leads to convolutions with Gaussian functions of variance equal the gate noise variance.
\end{widetext}

\section{Cluster state comparison cheat sheet}\label{app:C}
\begin{figure*}[h]
    \centering
    \includegraphics[width=\linewidth]{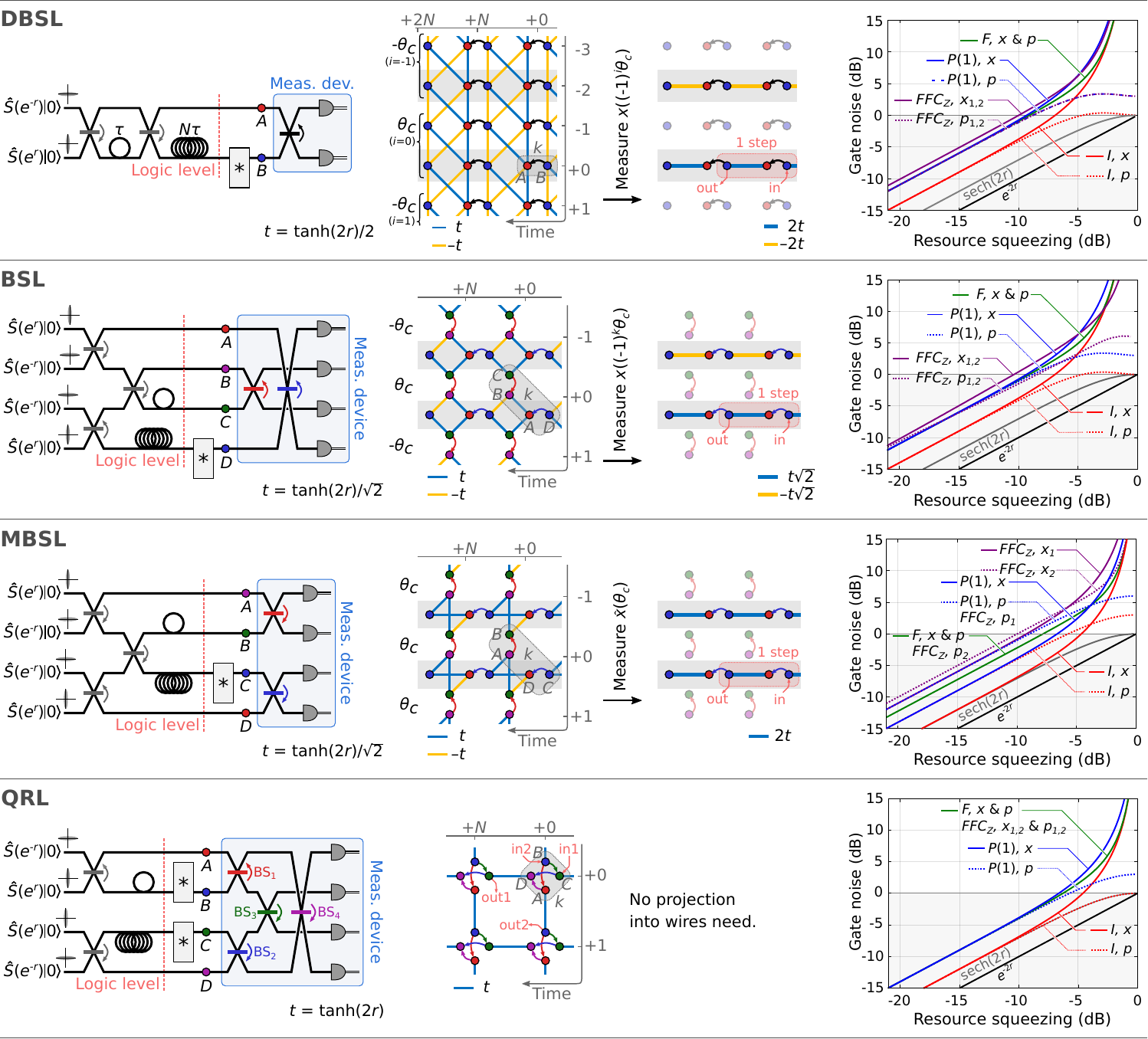}
    \caption{\label{fig:cheatsheet} Setup, cluster state in the logic level, and resulting gate noise of each the DBSL, BSL, MBSL, and QRL cluster states studied in the main text sections \ref{sec:3} and \ref{sec:4}. For the DBSL, BSL, and MBSL, the projection into wires is shown as well, while this is not required for computation on the QRL.}
\end{figure*}

In the main text section \ref{sec:3} and \ref{sec:4} the computation schemes on the different cluster states are analysed separately in order to facilitate easy lookup of a specific cluster state. The different schemes are discussed and compared in section \ref{sec:5}, while in this appendix the schemes are arranged side by side in Fig.~\ref{fig:cheatsheet} (located after the list of references) for easy comparison.

\end{document}